\documentclass[format=acmsmall, review=false, screen=true]{acmart}

\usepackage{tcolorbox}
\usepackage{graphicx}
\usepackage{array}
\usepackage{url}
\usepackage{fancybox}
\usepackage{multirow}
\usepackage{amssymb}
\usepackage{color}

\usepackage{colortbl}
\usepackage{balance}
\usepackage{fancyhdr}

\usepackage{subfig}
\usepackage{diagbox}
\usepackage{bbding}
\usepackage{placeins}

\usepackage{booktabs}
\usepackage{enumitem}
\usepackage{xcolor,lipsum}
\usepackage[linesnumbered, ruled,vlined]{algorithm2e}

\newcommand{\todo}[1]{}
\renewcommand{\todo}[1]{{\color{red} TODO: {#1}}}

\definecolor{light-gray}{rgb}{.906,  .902,  .902}

\newcommand{\major}[1]{\textcolor{black}{#1}}

\newcommand{\find}[1]{
\begin{tcolorbox}[leftrule=1mm,toprule=0mm,bottomrule=0mm,left=1pt,right=2pt,top=2pt,bottom=2pt] %[tile,size=fbox,boxsep=2mm,boxrule=0pt,top=0pt,bottom=0pt,borderline={0.5mm}{0pt}{black!70!white},colback=black!5!white]
#1
\end{tcolorbox}
}

\acmJournal{TOSEM}
\acmVolume{9}
\acmNumber{4}
\acmArticle{39}
\acmYear{2019}
\acmMonth{3}
\copyrightyear{2009}
\acmArticleSeq{9}

\setcopyright{acmlicensed}
\acmDOI{0000001.0000001}

\received{Mar 2019}
\received[revised]{?? 2019}
\received[accepted]{?? 2019}

\usepackage{bm}

\begin{document}
% Title portion. Note the short title for running heads
\title[]{Automating \textit{\major{TODO-missed} Methods} Detection and Patching}
% \titlenote{Corresponding Authors: Xin Xia}

\author{Zhipeng Gao}
%\orcid{0000-0002-2906-0598}
\affiliation{%
  \institution{Shanghai Institute for Advanced Study of Zhejiang University}
  \city{Shanghai,}
  % \state{VIC}
  % \postcode{3168}
  \country{China}
  }
\email{zhipeng.gao@zju.edu.cn}

\author{Yanqi Su}
\affiliation{%
  \institution{Australian National University}
  \city{Canberra,}
  % \state{VIC}
  % \postcode{3168}
  \country{Australia}
  }
\email{Yanqi.Su@anu.edu.au}

\author{Xing Hu}
\authornote{This is the corresponding author}
\affiliation{%
  \institution{Zhejiang University}
  \city{Hangzhou,}
  % \state{VIC}
  % \postcode{3168}
  \country{China}
  }
\email{xinghu@zju.edu.cn}

\author{Xin Xia}
\affiliation{%
  \institution{Huawei}
  \city{Hangzhou,}
  % \state{VIC}
  % \postcode{3168}
  \country{China}
  }
\email{xin.xia@acm.org}

% \author{David Lo}
% %\authornote{This is the corresponding author}
% \affiliation{%
%   \institution{Singapore Management University}
%   \city{Singapore,}
%   \country{Singapore}
%   }
% \email{davidlo@smu.edu.sg}

% \author{John Grundy}
% %\orcid{0000-0002-2906-0598}
% \affiliation{%
%   \institution{Monash University}
%   \city{Melbourne,}
%   \state{VIC}
%   \postcode{3168}
%   \country{Australia}
%   }
% \email{john.grundy@monash.edu}

\begin{abstract}
TODO comments are widely used by developers to remind themselves or others about incomplete tasks. 
In other words, TODO comments are usually associated with temporary or suboptimal solutions. 
In practice, all the equivalent suboptimal implementations should be updated (e.g., adding TODOs) simultaneously. 
However, due to various reasons (e.g., time constraints or carelessness), developers may forget or even are unaware of adding TODO comments to all necessary places, which results in the \textit{\major{TODO-missed} methods}. 
These ``hidden'' suboptimal implementations in \textit{\major{TODO-missed} methods} may hurt the software quality and maintainability in the long-term.
Therefore, in this paper, we propose the novel task of \textit{\major{TODO-missed} methods} detection and patching, and develop a novel model, namely {\sc TDPatcher} (\textbf{\underline{T}}O\textbf{\underline{D}}O-comment \textbf{\underline{Patcher}}), to automatically patch TODO comments to the \textit{\major{TODO-missed} methods} in software projects. 
Our model has two main stages: offline learning and online inference. 
During the offline learning stage, {\sc TDPatcher} employs the GraphCodeBERT and contrastive learning for encoding the TODO comment (natural language) and its suboptimal implementation (code fragment) into vector representations. 
For the online inference stage, we can identify the \textit{\major{TODO-missed} methods} and further determine their patching position by leveraging the offline trained model. 
We built our dataset by collecting \textit{TODO-introduced methods} from the top-10,000 Python GitHub repositories and evaluated {\sc TDPatcher} on them. 
Extensive experimental results show the promising performance of our model over a set of benchmarks. 
We further conduct an in-the-wild evaluation which successfully detects 26 \textit{\major{TODO-missed} methods} from 50 GitHub repositories.
\end{abstract}




\maketitle

\renewcommand{\shortauthors}{Zhipeng GAO et al.}
\renewcommand{\shorttitle}{Code Snippet Recommendation from Stack Overflow Posts}

\section{Introduction}
\label{sec:intro}
\begin{figure}
% \vspace{-7pt}
\centerline{\includegraphics[width=0.80\textwidth]{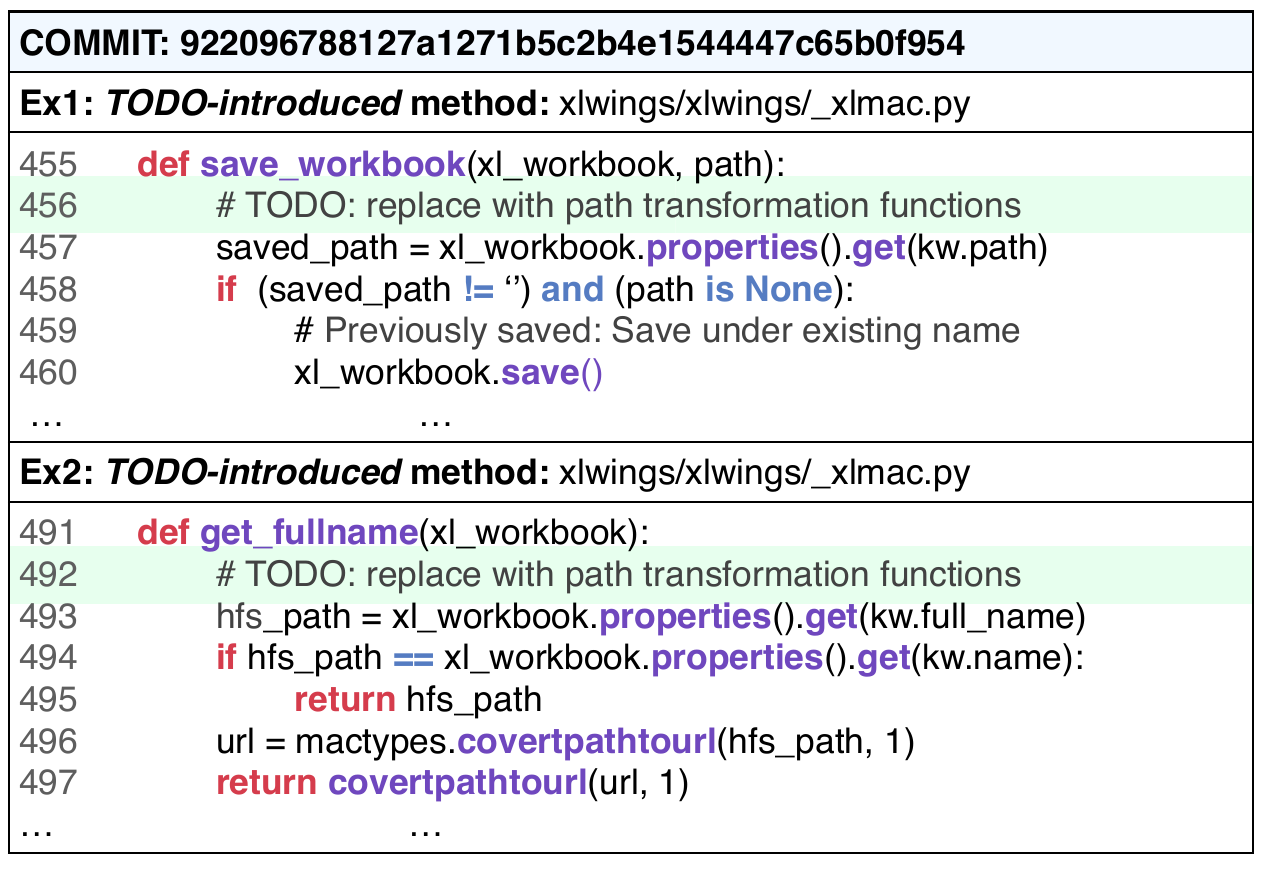}}
% \vspace*{-5pt}
\caption{Examples of TODO-introduced Methods}
\vspace{-10pt}
\label{fig:intro_example}
\end{figure}

TODO comments are widely used by developers to describe valuable code changes that can improve software quality and maintenance. 
In other words, the TODO comments pinpoint the current suboptimal solutions that developers should pay attention to in future development. 
For example, as shown in Example 1 of Fig.~\ref{fig:intro_example}, when a developer implemented the \texttt{save\_workbook} method, they added a TODO comment (Line 456, \textit{TODO: replace with path transformation functions}) to remind themselves or other developers for indicating the temporary path implementations (i.e., Line 457).
Ideally, once a TODO comment is added for a method, all the methods with equivalent suboptimal implementations should be updated (i.e., adding the same TODO comment) correspondingly. 
Example 2 in Fig.~\ref{fig:intro_example} shows such a situation, the method \texttt{get\_fullname} contains similar suboptimal implementations (i.e., Line 493) as method \texttt{save\_workbook}. 
Therefore, the developers added a TODO comment to the above two methods simultaneously in one commit. 
However, due to the complexity of the ever-increasing software systems (e.g., hundreds of files with thousands of methods) and frequent refactoring of the source code (e.g., taking over code from others), developers may forget or even are unaware of such suboptimal implementations somewhere else in the project, resulting in \textit{\major{TODO-missed} methods}.

In our work, the definition of \textit{\major{TODO-missed} method} is as follows: 
given a method pair $\langle  \mathbf{m}_{1}, \mathbf{m}_{2} \rangle$ which has equivalent suboptimal implementations $\mathbf{I}$, if $\mathbf{m}_{1}$ was added with a TODO comment $\mathbf{T}$ to indicate $\mathbf{I}$ but \major{$\mathbf{m}_{2}$} was not, we then define 
$\mathbf{m}_{2}$ as a \textbf{\textit{\major{TODO-missed} method}}, and $\mathbf{m}_{1}$ as a \textbf{\textit{TODO-introduced method}}. 
Methods with proper TODO comments can help developers understand the source code context and prevent introducing mistakes when touching unfamiliar code. 
On the contrary, the \textit{\major{TODO-missed} methods} miss the TODO comment specification, resulting in the suboptimal implementations being ignored (for a long time) or never getting revisited unless causing any damages. 
Take the above methods as an example, if the TODO comment is not specified clearly in \texttt{get\_fullname}, the developers can easily forget this unfinished task, the problem is even worsened as software constantly evolves and developers frequently join and leave the development team. 
It is thus helpful to notify developers of \textit{\major{TODO-missed} methods} in software projects before any unwanted side effects are caused.

In this paper, we aim to identify \textit{\major{TODO-missed} methods} when new TODO comments are added. 
Manually checking \textit{\major{TODO-missed} methods} is a time-consuming and error-prone process, this is especially true for large software projects, usually with hundreds of files and thousands of lines of code. 
It would cost a great amount of effort if developers inspect all methods line by line to identify \textit{\major{TODO-missed} methods}.  
Therefore, it is beneficial and desirable to have a tool for automating the checking of \textit{\major{TODO-missed} methods} in practice. 
To perform this task, the following two key issues need to be handled properly: 

\begin{enumerate}
    \item \textbf{\textit{\major{TODO-missed} methods} detection.} Identifying \textit{\major{TODO-missed} methods} first requires understanding the semantics of the TODO comment and its suboptimal implementation and the relationship between them. The comment and source code are of different types (i.e., free-form natural languages and formal programming languages), which causes naturally a gap between them. As shown in Fig.~\ref{fig:intro_example} examples, TODO comments and their suboptimal implementations are semantically related but lexically isolated. 
    As a result, it is a non-trivial task to determine whether a method should include a particular TODO comment or not. 
    Therefore, an effective tool for detecting \textit{\major{TODO-missed} methods} requires capturing the semantic mapping and implicit connections between the TODO comment and its corresponding suboptimal implementation(s). 
    \item \textbf{\textit{\major{TODO-missed} methods} patching.} 
    After a \textit{\major{TODO-missed} method} is detected, our next step aims to locate the exact suboptimal positions (e.g., the source code lines) that the TODO comment should be patched to, namely \textit{\major{TODO-missed} methods} patching. 
    Patching TODO comments to the right location is also a non-trivial task with respect to the following reasons: it is very difficult, if not possible, to determine where a TODO comment should be added by just reading the method source code. 
    For example, as shown in Example 2 in Fig.~\ref{fig:intro_example}, multiple statements within \texttt{get\_fullname} method are associated with ``path'' actions, one cannot easily claim which code line is connected to the TODO comment. A more intuitive and reliable way is to take the \textit{TODO-introduced method} as a reference and check the equivalent suboptimal positions by pairwise comparison. 
\end{enumerate} 

To alleviate the above issues and help developers better maintain TODO comments, in this work, we propose a novel framework named {\sc TDPatcher} (\textbf{\underline{T}}O\textbf{\underline{D}}O-comment \textbf{\underline{Patcher}}) to automatically patch the TODO comments to the \textit{\major{TODO-missed} methods}. 
The main idea of our approach is two folds: 
(i) \textbf{Code Embeddings}: code patterns (e.g., the suboptimal implementations), as well as code documentation (e.g., TODO comments), can be automatically encoded into contextualized semantic vectors via the techniques adapted from the large-scale pre-trained models (e.g., GraphCodeBERT). 
(ii) \textbf{Contrastive Learning}: learning semantic representation for an individual method alone is not sufficient, 
it is necessary to explore the correlations and implicit differences between the \textit{TODO-introduced methods} and \textit{\major{TODO-missed} methods}. 
In this work, we adopt the contrastive learning strategy to teach the model to pull positive samples together (i.e., methods with equivalent suboptimal implementations) while simultaneously pushing negative samples (i.e., irrelevant methods) apart.

{\sc TDPatcher} consists of two stages: offline learning and online inference. 
During offline learning, we collect \textit{TODO-introduced methods} (a method which newly adds TODO comment) from the Top-10,000 Python GitHub repositories.
We then automatically establish $\langle anchor, positive, negative \rangle$ triplet training samples in terms of whether the \textit{TODO-introduced method} can be paired. 
{\sc TDPatcher} first employs the GraphCodeBERT~\cite{guo2020graphcodebert} model to embed $anchor$, $positive$ and $negative$ samples into contextualized vectors respectively. 
{\sc TDPatcher} then applies contrastive learning strategy~\cite{hadsell2006dimensionality, gao2021simcse} to further learn the discriminative vector representations. 
The goal of our framework is that equivalent suboptimal implementations (i.e., positive pairs) should be as close as possible in hidden vector space, while the irrelevant code patterns (i.e., negative pairs) should be as far away as possible in the space. 
When it comes to online prediction, for a given \textit{TODO-introduced method}, {\sc TDPatcher} first transforms its suboptimal code implementations into vectors, then locates the code fragments that have the nearest vectors to the suboptimal implementations.

To verify the effectiveness of {TDPatcher}, we conducted extensive experiments on the Python dataset. 
By comparing with several benchmarks, the superiority of our proposed {\sc TDPatcher} model is demonstrated.
In summary, this work makes the following main contributions: 
\begin{enumerate}
    \item We propose a novel task of \textit{\major{TODO-missed} methods} detection and patching, we build a large dataset for checking \textit{\major{TODO-missed} methods} from top-10,000 Python Github repositories. To the best of our knowledge, it is the first large dataset for this task. 
    
    \item We propose a novel model, {\sc TDPatcher}, to automatically detect and patch \textit{\major{TODO-missed} methods} when TODO comments are introduced. 
    {\sc TDPatcher} can help developers to increase the quality and maintainability of software, and alleviate the error-prone code review process. 
    
    \item We extensively evaluate {\sc TDPatcher} using real-world popular open-source projects in Github. {\sc TDPatcher} is shown to outperform several baselines and reduce the developer's efforts in maintaining the TODO comments.  
    
    \item We have released our replication package, including the dataset and the source code of {\sc TDPatcher}~\cite{tdpatcher}, to facilitate other researchers and practitioners to repeat our work and verify their ideas. 
    
\end{enumerate}

The rest of the paper is organized as follows.
Section~\ref{sec:motiv} presents the motivating examples and user scenarios of our study. 
Section~\ref{sec:approach} presents the details of our approach.
Section~\ref{sec:data} presents the data preparation for our approach. 
Section~\ref{sec:eval} presents %the baseline methods, the evaluation metrics, and 
the evaluation results.
Section~\ref{sec:disc} presents the in-the-wild evaluation.
Section~\ref{sec:threats} presents the threats to validity. 
Section~\ref{sec:related} presents the related work.
Section~\ref{sec:con} concludes the paper with possible future work.

\section{Motivation}
\label{sec:motiv}

\begin{figure}
% \vspace{-3pt}
\centerline{\includegraphics[width=0.95\textwidth]{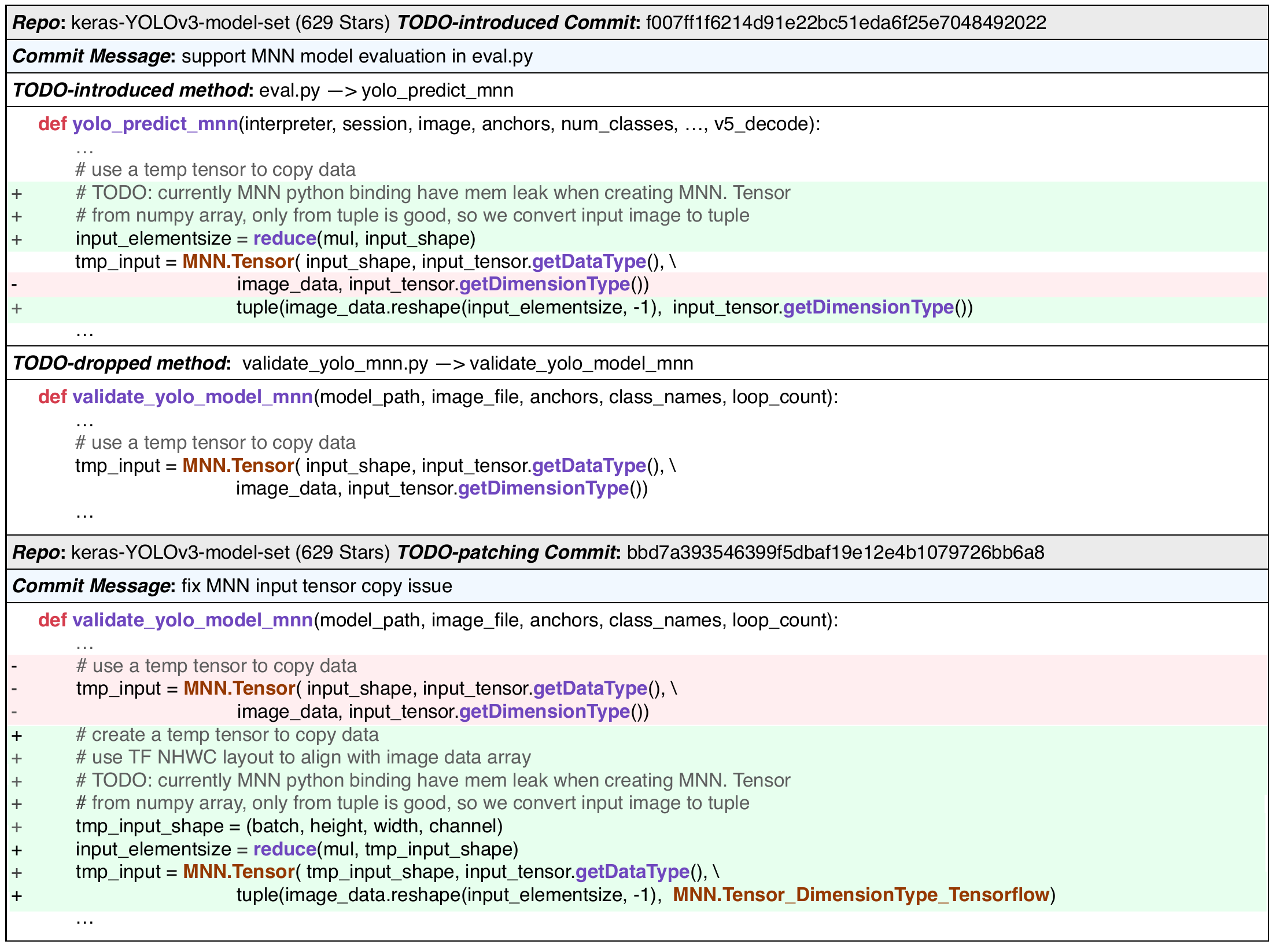}}
\vspace*{-5pt}
\caption{\major{Motivating Example 1}}
\label{fig:moti_example1}
% \vspace{-9pt}
\end{figure}

\major{
In this section, we first show the motivating examples from real-world \textit{\major{TODO-missed} methods}. 
We then present user scenarios that employ our proposed {\sc TDPatcher} to address these problems.
}

\subsection{Motivating Examples}

\begin{figure}
% \vspace{-3pt}
\centerline{\includegraphics[width=0.95\textwidth]{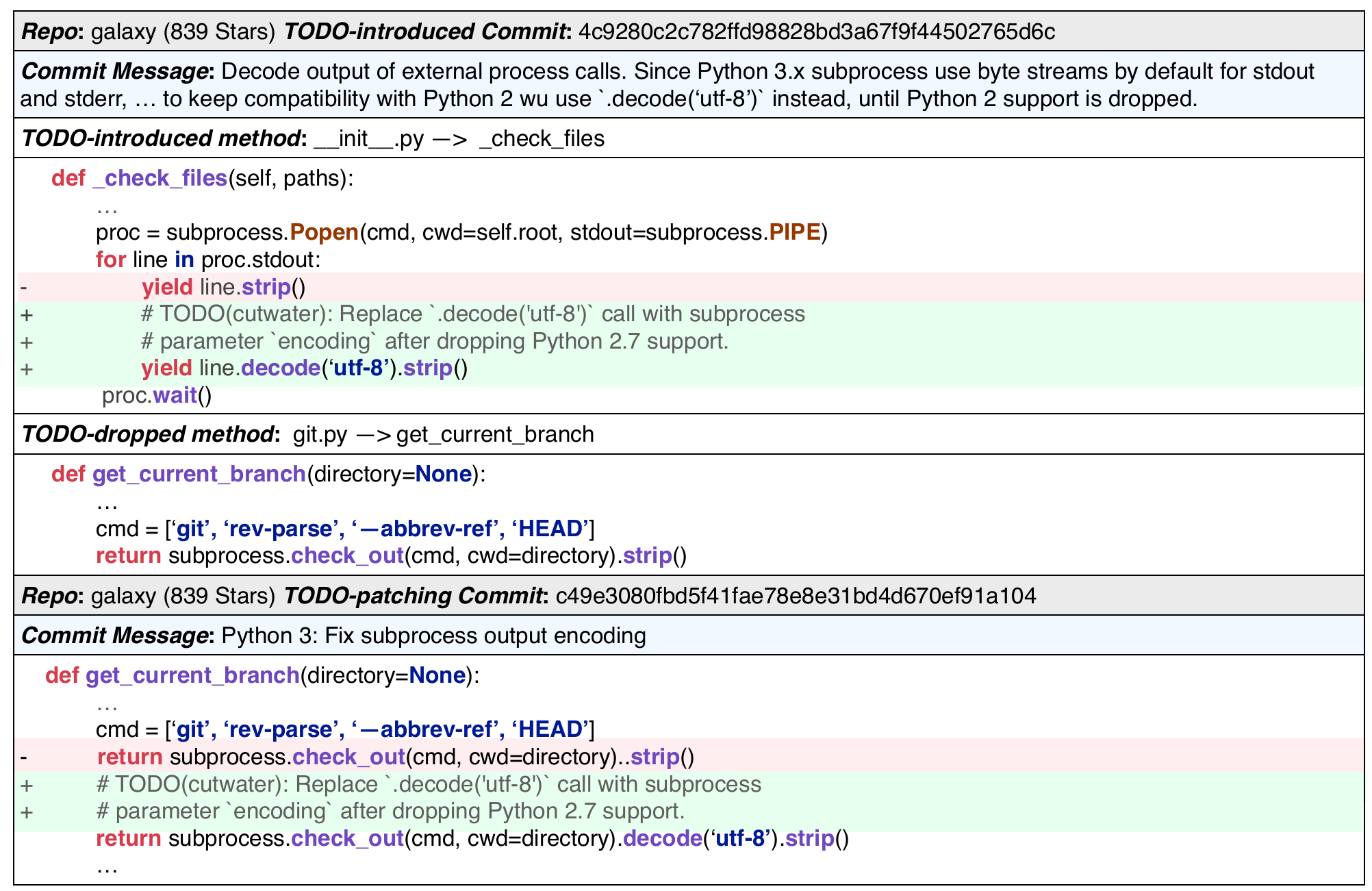}}
\vspace*{-5pt}
\caption{\major{Motivating Example 2}}
\label{fig:moti_example2}
% \vspace{-9pt}
\end{figure}

\major{Due to the large scale of modern software projects, developers may forget or even be unaware of adding TODO comments to the methods that are supposed to, leading to the introduction of \textit{\major{TODO-missed} methods}. 
Fig.~\ref{fig:moti_example1} and Fig.~\ref{fig:moti_example2} show two examples of \textit{\major{TODO-missed} methods} in real-world GitHub repositories. 
From these examples we can see that: 
\begin{enumerate}
    \item \textbf{\textit{\major{TODO-missed} methods} can decrease the software quality and maintainability and induce potential problems and/or bugs in subsequent development.} 
    An example is shown in Fig.~\ref{fig:moti_example1}, the developer added a TODO comment to \texttt{yolo\_predict\_mnn} method, noting that ``\textit{TODO: currently MNN python binding have mem leak when creating MNN, ... so we convert input image to tuple}''.  
    To address this memory leak issue, the developer modified the \texttt{image\_data} to a tuple. 
    Unfortunately, the developer neglected the same suboptimal implementation in another method (i.e., \texttt{validate\_yolo\_model\_mnn}), leading to the latter becoming a \textit{TODO-missed method}. 
    As a result, the same problem could be caused when the \texttt{validate\_yolo\_model\_mnn} method is triggered. 
    Upon manual inspection of the project's commit history, we found that the developer eventually added the missing TODO comment back to \texttt{validate\_yolo\_model\_mnn} and fixed the problem within the \textit{TODO-missed method} using the same way as the \textit{TODO-introduced method}. 
    The commit message (``\textit{fix MNN input tensor copy issue}'') further confirms our assumption that the problem was caused by overlooking the suboptimal implementation within the \textit{TODO-missed method.} 
    \item \textbf{Identifying \textit{TODO-missed methods} is a non-trivial task, given diverse forms of equivalent suboptimal implementations.} 
    Fig.~\ref{fig:moti_example2} shows another example of \textit{TODO-missed method} in the \texttt{galaxy} project. 
    As can be seen, the \textit{TODO-introduced method} (i.e., \texttt{\_check\_files}) and the \textit{TODO-missed method} (i.e., \texttt{get\_current\_branch}) have the same encoding problem related to the \texttt{subprocess} output. 
    Despite sharing the same problem, the suboptimal code patterns within the \textit{TODO-introduced method} (i.e., \texttt{yield line.strip()}) and the \textit{TODO-missed method} (i.e., \texttt{return subprocess.check\_out().strip()}) differ significantly. 
    The developer can easily ignore such suboptimal code implementations due to time constraints and/or carelessness, and has to spend extra time and effort for debugging when problems are eventually exposed. 
    Regarding this particular case, the developer fixed the encoding issue in the \textit{TODO-missed method} after a week when the TODO comment was initially added. 
    During this time, other team members may revise the code without noticing the suboptimal code patterns in the source code, therefore any code refactoring related to this \textit{TODO-missed method} can be influenced, posing potential threats to software maintenance and evolution. 
\end{enumerate}
In summary, the \textit{TODO-missed methods} can be easily ignored by developers and introduce potential problems and/or bugs in subsequent development. 
Therefore, we propose {\sc TDPatcher} in this work, which aims to make the hidden suboptimal implementations \textit{observable} by adding TODO comments to the \textit{\major{TODO-missed} methods}. 
With the help of {\sc TDPatcher}, developers will be reminded of the ignored suboptimal implementations just-in-time, such immediate feedback ensures the context is still fresh in the minds of developers. 
This fresh context can help developers to be aware of their suboptimal code and thus avoid the introduction of the \textit{TODO-missed methods}. 
}

\subsection{User Scenarios}
We illustrate a usage scenario of {\sc TDPatcher} as follows:

\textbf{Without Our Tool:}
Consider Bob is a developer. 
One day, when Bob implemented a method to add a new feature, he realized that the current solution was suboptimal. 
He then adds a TODO comment to indicate the features that are currently not supported or the optimizations that need to be implemented. 
Besides the current method, there are multiple methods with the equivalent suboptimal implementation that are supposed to be handled correspondingly. 
However, due to time constraints or unfamiliarity with the software system, Bob is not aware of other suboptimal positions and thus introduces the \textit{\major{TODO-missed} methods} unconsciously. 
These \textit{\major{TODO-missed} methods} may negatively impact program quality and maintenance. 
Furthermore, when other developers take over the code from Bob, they have no idea that the method is suboptimal, the new updates on this method are risky and may introduce bugs in the future. 

\textbf{With Our Tool:}
Now consider Bob adopts our tool, {\sc TDPatcher}. 
After introducing TODO comments and their suboptimal solutions, Bob can use our tool to automatically identify the presence of the \textit{\major{TODO-missed} methods}, such immediate feedback can ensure the code context is still fresh in the developer's mind. 
Moreover, {\sc TDPatcher} can locate the specific source code line where the TODO comments are supposed to be added, thus helping developers to locate the suboptimal positions with less inspection effort. 
With the help of our tool, Bob successfully finds and patches the \textit{\major{TODO-missed} methods} in the current software repository efficiently, which increases the reliability and maintenance of the system and decreases the likelihood of introducing bugs. 

\section{Approach}
\label{sec:approach}
In this section, we first define the task of \textit{\major{TODO-missed} methods} detection and patching for our study. 
\major{
We then present the details of our data preparation and proposed approach. 
}
\subsection{Task Definition}
Our tool aims to automatically detect and patch \textit{\major{TODO-missed} methods} in software projects. 
Formally, given a method pair $\langle \mathbf{m_{1}}, \mathbf{m_{2}} \rangle$, let $\mathbf{m_1}$ be the \textit{TODO-introduced method} and $\mathbf{T}$ be the TODO comment associated with $\mathbf{m_1}$, let $\mathbf{m_2}$ be the candidate method to be checked. 
Our first task is to find a function \texttt{detect} so that: 
\begin{equation}
\label{eq:detect}
detect ( \mathbf{m_{1}}, \mathbf{m_{2}}, \mathbf{T} ) =   
\left\{ \begin{array}{rcl}
1 & \mbox{for} & \mathbf{m_{1}} \rightleftharpoons \mathbf{m_{2}} \\
0 & \mbox{for} & otherwise
\end{array}\right. 
\end{equation}
where $\mathbf{m_{1}} \rightleftharpoons \mathbf{m_{2}}$ denotes that the $m_1$ and $m_2$ contains the equivalent suboptimal implementations. 
Our second task is to add the TODO comment $\mathbf{T}$ to the right position of $\mathbf{m_{2}}$ if $detect(\mathbf{m_{1}}, \mathbf{m_{2}}, \mathbf{T})=1$. 
Let $\mathbf{p}$ be the desired adding position of $\mathbf{T}$ in $\mathbf{m_{2}}$. 
\major{Our} second task is to find a function \texttt{patch} so that: 

\begin{equation}
\label{eq:patch}
patch ( \mathbf{m_{1}}, \mathbf{m_{2}}, \mathbf{T} ) = \mathbf{p}
\end{equation}

% {\sc TDPatcher} leverage the code embedding and contrastive learning techniques to approximate the \texttt{detect} and \texttt{patch} simultaneously. 
% Particularly, {\sc TDPatcher} first uses the large-scale pre-trained model to jointly embeds code patterns (i.e., suboptimal implementations) and documentations (i.e., TODO comments) into a unified vector space. 
% Following that {\sc TDPatcher} apply the contrastive learning strategy to further model the discriminative features, in such a way that equivalent code implementations will be embedded into nearby vectors and can be matched by vector similarities. 

\major{
\subsection{Data Collection}
We first present the details of our data collection process.
We build our dataset by collecting data from the top 10,000 Python repositories (ordered by the number of stars) in GitHub, up until December 25th, 2022. 
We have to mention that while this study utilizes  Python for investigation, our approach could be applied to source code written in any programming language.
}

\major{
\subsubsection{Identifying TODO-introducing Commits}
\label{subsec:identify-todo}
We first collected and cloned the top-10K Python repositories from GitHub until December 2022. 
The git repository stores software update history, each update comprises a \texttt{diff} file that shows the differences between the current and previous versions of the affected files. 
For each cloned repository, we first checkout all the commits from the repository history. 
Following that, for each commit, if ``TODO'' appears within \textbf{added lines} of the \texttt{diff}, we consider this commit as a \emph{TODO-introducing commit}. 
Among the top-10K Python repositories, we discovered that 6,519 of them contained \emph{TODO-introducing commits}.
Finally, we have collected 148,675 \textit{TODO-introducing commits} (including 196,945 TODOs since there are commits containing more than one TODO) from these 6,519 Python repositories. 
Considering the forks of an original repository could introduce noise data samples and contaminate the training and testing dateset, we further checked that all 6,519 Python repositories are original repositories and not forks from other repositories. 
\subsubsection{Cleaning TODO-introducing Commits.}
\label{subsec:cleaning-todo}
Although we have gathered \textit{TODO-introducing commits} from Python repositories, it is not guaranteed that all TODO comments are added to Python methods. 
In this step, we only preserve \emph{TODO-introducing commits} if they contain TODO comments added to Python source files and remove any other unrelated commits. 
To achieve this, we apply \texttt{Pydriller}~\cite{Spadini2018} (a framework for analyzing Git repositories), to extract modified files of each \emph{TODO-introducing commit}, and only retain commits when their associated modified files are Python source files (ending up with \texttt{.py}). 
As a result, we retain a total of 112,951 verified \emph{TODO-introducing commits} (including 152,724 TODOs) from 5,888 Python repositories.  
\subsubsection{Extracting TODO-introduced Methods} 
\label{subsec:extracting-todo}
In this step, We extract methods that are associated with the TODO comments. 
Particularly, for a given verified \emph{TODO-introducing commit}, we first check out affected files from this commit.  
We then pinpoint the line number of the TODO comment and automatically extract the \emph{TODO-introduced method} if the method's code range covers the line of the TODO comment. 
During this process, we apply the following rules: 
\textbf{(i) Rule1: The TODO comments that consist of less than 3 words are excluded.} 
These TODOs are often too short to carry meaningful code indications. 
As a result, we removed 9,660 \emph{TODO-introducing commits} (including 15,957 TODOs), leaving us with 136,767 TODOs for this step.
\textbf{ (ii) Rule2: The TODO comments that are not located within a method are excluded.} 
Since our focus is on addressing \textit{TODO-missed methods} in this study, we disregard TODO comments that are not associated with any methods. 
By doing so, we removed 39,354 TODOs that were outside of methods and retained 97,413 within method TODOs. 
\textbf{(iii) Rule3: The methods that failed to parse are excluded.} 
To extract method-level information for downstream tasks, we use Python 2.7/3.7 standard library \texttt{ast} to produce the file-level Abstract Syntax Tree (AST) for Python 2 and Python 3 projects respectively. 
We then extract every individual method and class method source code via inspecting the method related AST node (e.g., the \texttt{FunctionDef} and \texttt{AsyncFunctionDef}). 
We use the \texttt{autopep8} to overcome the issues of different styles and white space or tab conventions. 
Finally, 1,451 methods failed to parse due to syntax error, 95,962 \emph{TODO-introduced methods} are successfully parsed and extracted for downstream tasks. 
\subsubsection{Grouping TODO-introduced Methods} 
\label{subsec:group-todo}
This step is responsible for grouping methods with the same TODO comments together.  
Different methods may contain the same TODO comments, which indicates the same suboptimal code implementations. 
In this step, for a given project, we automatically check each \textit{TODO-introduced method} within this project. 
Methods with the same TODO comment will be grouped together, while methods with unique TODO comment will be excluded. 
As a result, each TODO comment will be paired with a set of methods (e.g., different methods or the same method of different TODO locations). 
Finally, the \textit{TODO-introduced method} groups, $\mathcal{G} = \{ (\mathbf{T_{i}}, \mathbf{m_{i_{1}}}, \mathbf{m_{i_{2}}}, ..., \mathbf{m_{i_{n}}}) \}_{i=1}^{g}$ are constructed, where $\mathbf{m_{i_{1}}}, ..., \mathbf{m_{i_{n}}}$ are a group of methods with the same TODO comment $\mathbf{T_{i}}$. 
Each group of methods is regarded as methods with equivalent suboptimal code implementations. 
Among the 95,962 \textit{TODO-introduced methods}, 18,126 of them are grouped into 6,855 \textit{TODO-introduced method} groups, while the other 77,386 \textit{TODO-introduced methods} are excluded. 
% Finally, 18,126 \textit{TODO-introduced methods} are 
% we obtain 6,855 \textit{TODO-introduced} method groups in total. 
\subsubsection{Manual Validation}
\label{subsec:manual-validation}
Since we automatically built our dataset of \textit{TODO-introduced methods} from Top-10K Python repositories, we can not ensure that there are no outlier cases, or noisy data, during the data preparation process.  
Therefore, we further performed a manual checking step to validate the quality of our constructed dataset. 
Specifically, we randomly sampled 100 \textit{TODO-introduced method} groups from our dataset. 
Then the first author manually examined whether \textit{TODO-introduced methods} within each group contains equivalent suboptimal implementations. 
Finally, 96 groups of methods are marked as methods with equivalent suboptimal implementations. 
Thus, we are confident with the quality of our constructed dataset. 
}

\subsubsection{Empirical Findings.}
\major{
We have gathered the following insights from our data collection process: 
(i) \textbf{TODOs are widely used by developers in open-source projects.} 
As shown in Section~\ref{subsec:identify-todo}, among the top-10K Python repositories, 6,519 of them incorporate TODOs during the software development. 
This confirms the significant role of TODO comments in managing and coordinating diverse programming tasks. 
(ii) \textbf{The majority of TODOs are added within methods.} 
As demonstrated in our preliminary investigation in Section~\ref{subsec:extracting-todo}, among the 136,767 introduced TODOs, approximately 71\% (97,413 TODOs) are inserted within methods, which verifies the importance and necessity of targeting our objectives, i.e.,\textit{TODO-missed methods}. 
(iii) We have identified 6,855 \textit{TODO-introduced method} groups in Section~\ref{subsec:group-todo}.  
Ideally, methods within the same group should be updated (i.e., adding TODOs) simultaneously. 
However, after our investigation, 3,436 groups (i.e., 50.1\%) of methods added TODOs within the same commit, while the remaining 3,419 (i.e., 49.9\%) groups added TODOs across different commits.
This phenomenon signals that \textbf{a large number of TODOs are not added in time by developers, leading to the introduction of \textit{TODO-missed methods}.} 
It is thus highly desirable to have a tool that provides just-in-time automatic detection of \textit{TODO-missed methods} and add them before they mislead developers and/or cause any unwanted side effects. 
This automated process of detecting and patching \textit{TODO-missed methods} can substantially increase the quality and reliability of the software system.  
}

\subsection{Approach Details}
\begin{figure*}
% \vspace{-7pt}
\centerline{\includegraphics[width=0.95\textwidth]{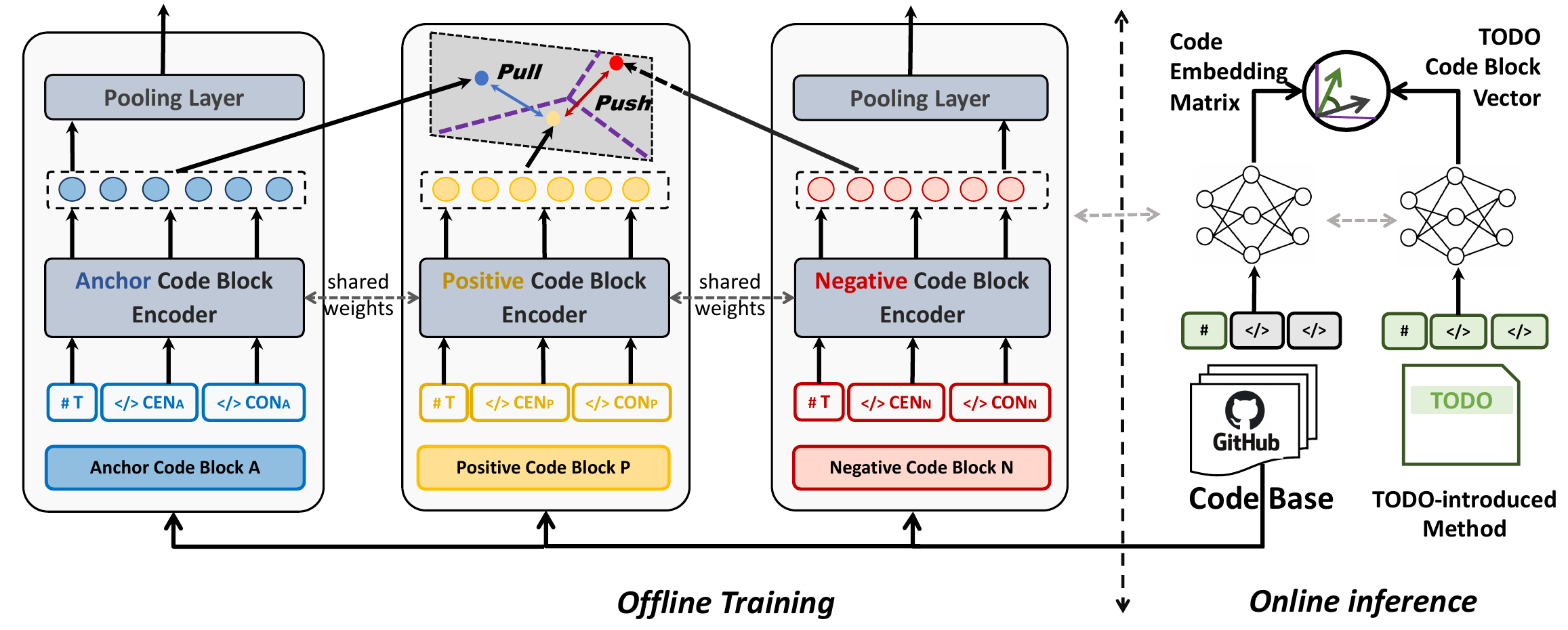}}
\vspace*{-5pt}
\caption{\major{Overall Framework of Our {\sc TDPatcher}}} 
% \zp{To be Updated}} 
\label{fig:workflow}
% \vspace{-8pt}
\end{figure*}

\major{
In this section, we first present the details of constructing triplets model inputs, as shown in Fig.~\ref{fig:code_block_example}. 
We then describe the overall framework of our approach, named {\sc TDPatcher}, as illustrated in Fig.~\ref{fig:workflow}. 
The approach details are as follows: 
}

\major{
\subsubsection{Model Inputs Construction.}
\label{subsec:model_inputs}
To capture semantic correlations between equivalent code implementations, we adopt the idea of contrastive learning. 
Different from traditional software engineering tasks such as defect prediction~\cite{tantithamthavorn2016automated, wang2016automatically}, contrastive learning requires triplets as inputs~\cite{wei2022clear, gao2021simcse, chen2021cil}. 
Therefore, we need to prepare the data as triplets for model inputs. 
Each triplet is composed of three code blocks, namely \textbf{anchor code block} (denoted as $\mathbf{A}$), \textbf{positive code block} (denoted as $\mathbf{P}$), and \textbf{negative code block} (denoted as $\mathbf{N}$) respectively.  
Each code block consists of three parts: $\mathbf{T}$, $\mathbf{CEN}$ and $\mathbf{CON}$, where $\mathbf{T}$ represents the \textit{TODO comment}, $\mathbf{CEN}$ represents the \textit{TODO centrepiece}, and $\mathbf{CON}$ represents the \textit{TODO context} respectively. 
% Here, we introduce the details of extracting code blocks from methods. 
Fig.~\ref{fig:code_block_example} demonstrates the code block generation process. 
The detailed definitions of \textit{TODO centrepiece} and \textit{TODO context} are explained as follows:
% We introduce the detailed definition 
\begin{itemize}
    \item \textbf{Anchor Code Block Generation.} After data collection, we obtain the \textit{TODO-introduced method} groups $\mathcal{G}$, for a given group $(\mathbf{T}, \mathbf{m_1}, \mathbf{m_2}, ..., \mathbf{m_n})$ in $\mathcal{G}$, we randomly select a method $\mathbf{m_{k}} (1 \leq k \leq n)$ as the anchor method. We construct the anchor code block for $\mathbf{m_{k}}$ as follows: we identify the code line following the TODO comment as the \textit{TODO centrepiece}; We consider the two code lines before and after the TODO comment (excluding the \textit{TODO centrepiece}) as the \textit{TODO context}. 
    Finally, the TODO comment, the \textit{TODO centrepiece}, and the \textit{TODO context} make an anchor code block. In summary, the anchor code block $\mathbf{A} = [\mathbf{T}, \mathbf{CEN_{A}}, \mathbf{CON_{A}}]$, where $\mathbf{CEN_{A}}$ and $\mathbf{CON_{A}}$ are the \textit{TODO centrepiece} and \textit{TODO context} of the anchor method. 
    \item \textbf{Positive Code Block Generation.} 
    The positive code block refers to the code fragment that shares equivalent suboptimal implementations with the anchor code block. 
    For a given \textit{TODO-introduced method} group $(\mathbf{T}, \mathbf{m_1}, ..., \mathbf{m_n})$, once the method $\mathbf{m_k}$ is selected as the anchor method, each of the rest methods $\mathbf{m_j} (1 \leq j \leq n, j \neq k)$ will serve as a candidate positive method. 
    Given the positive method $\mathbf{m_j}$, we generate the positive code block exactly the same as making the anchor code block. 
    Similarly, the positive code block $\mathbf{P} = [\mathbf{T}, \mathbf{CEN_{P}}, \mathbf{CON_{P}}]$ can be constructed, where $\mathbf{CEN_{P}}$ and  $\mathbf{CON_{P}}$ are the \textit{TODO centrepiece} and \textit{TODO context} of the positive method $\mathbf{m_j}$. 
    \item \noindent\textbf{Negative Code Block Generation.} The negative code block refers to the irrelevant code fragment in terms of the anchor code block. 
    The negative code block also includes three segments. 
    The TODO comment is the same comment with the anchor/positive code block. Regarding the \textit{TODO centrepiece}, we randomly select a source code line from a non-TODO method as the negative \textit{TODO centrepiece}, the two code lines before and after the centrepiece are regarded as the negative \textit{TODO context}. 
    Likewise, a negative code block $\mathbf{N} = [\mathbf{T}, \mathbf{CEN_{N}}, \mathbf{CON_{N}}]$, where $\mathbf{CEN_{N}}$ and $\mathbf{CON_{N}}$ represents the negative TODO centrepiece and negative TODO context respectively. 
\end{itemize} 
\begin{figure*}
\vspace{-7pt}
\centerline{\includegraphics[width=0.99\textwidth]{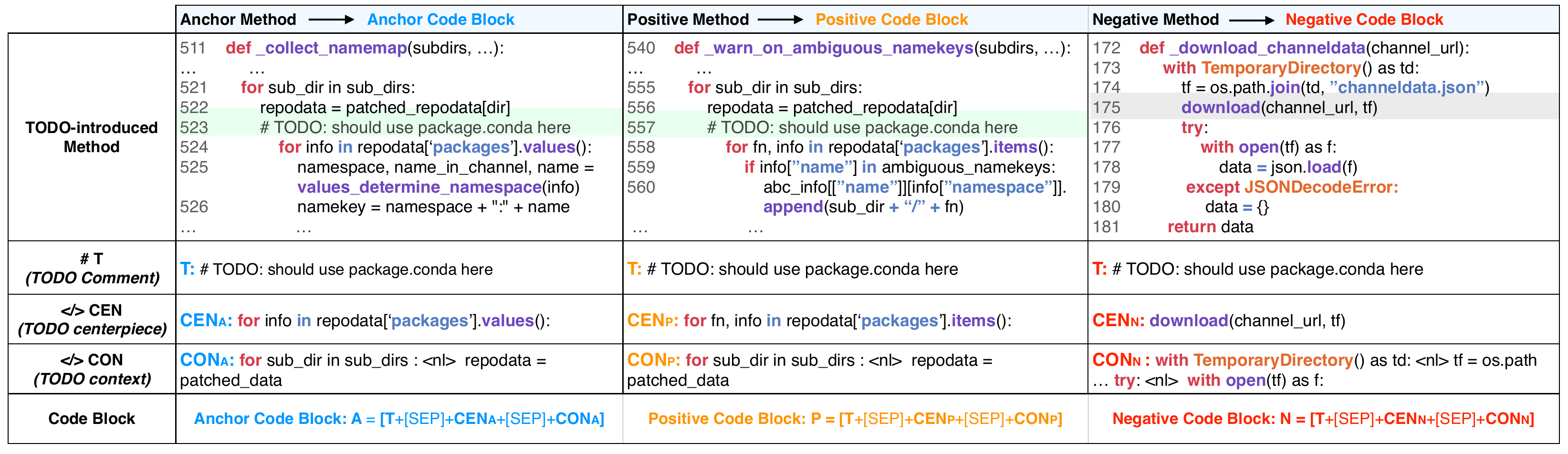}}
\vspace*{-3pt}
\caption{\major{Code Block Generation Examples}}
\label{fig:code_block_example}
\vspace{-8pt}
\end{figure*}
So far, given a \textit{TODO-introduced method} group $(\mathbf{T}, \mathbf{m_1}, \mathbf{m_2}, ..., \mathbf{m_n})$, once the anchor method is determined, the remaining method within the group is considered a positive method. 
Then each positive method will be paired with the anchor method and a negative method to establish a triplet sample. 
In other words, the \textit{TODO-introduced method} group $(\mathbf{T}, \mathbf{m_1}, \mathbf{m_2}, ..., \mathbf{m_n})$ can generate $n-1$ triplet samples. 
From each triplet sample, the anchor code block $\mathbf{A}$, positive code block $\mathbf{P}$ and negative code block $\mathbf{N}$ are extracted automatically for model inputs, denoted as $\langle \mathbf{A}, \mathbf{P}, \mathbf{N} \rangle$. 
The positive code block $\mathbf{P}$ paired with the anchor code block $\mathbf{A}$ makes a positive pair (i.e., $\langle \mathbf{A}, \mathbf{P} \rangle$), while the negative code block $\mathbf{N}$ paired with the anchor code block $\mathbf{A}$ makes a negative pair (i.e., $\langle \mathbf{A}, \mathbf{N} \rangle$). 
In this work, the positive pairs are code blocks with equivalent suboptimal implementations, while the negative pairs are code blocks with irrelevant implementations. 
Since our code block consists of three parts (i.e., \textit{TODO comment}, \textit{TODO centrepiece} and \textit{TODO context}), we add a special token \texttt{[SEP]} between different parts to further separate the natural language (TODO comment) and code implementations (i.e., TODO centrepiece and TODO context), which has been proven to be effective for bridging the gap between heterogeneous data. 
Finally, we constructed 10,775 triplet samples from  6,855 \textit{TODO-introduced method groups.}
% We can pair each anchor method $\mathbf{m_a}$ with a positive method $\mathbf{m_p}$, and a negative method $\mathbf{m_n}$ to establish a triplet sample (i.e., $\langle \mathbf{m_a}, \mathbf{m_p}, \mathbf{m_n} \rangle$). 
% we have constructed the triplet inputs for our model, every single input is a combination of three code blocks, denoted as $\langle \mathbf{A}, \mathbf{P}, \mathbf{N} \rangle$. 
% The positive code block $\mathbf{P}$ paired with the anchor code block $\mathbf{A}$ makes a positive pair (i.e., $\langle \mathbf{A}, \mathbf{P} \rangle$), while the negative code block $\mathbf{N}$ paired with the anchor code block $\mathbf{A}$ makes a negative pair (i.e., $\langle \mathbf{A}, \mathbf{N} \rangle$). 
% In this work, we regard the code blocks with equivalent suboptimal implementations as positive pairs and code blocks with irrelevant implementations as negative pairs. 
}

\subsubsection{Code Embedding Networks}
\major{
Due to the lexical gap between the TODO comments (i.e., natural language) and their suboptimal implementations (i.e., source code), it is difficult for traditional embedding techniques such as Bag-of-words, TF-IDF, and word2vec to embed both source code and natural language as vectors. 
In order to capture the semantic correlation between the TODO comment and its suboptimal solutions, we employ the large-scale pretrained model, GraphCodeBERT~\cite{guo2020graphcodebert}, as the encoder component for our model. 
GraphCodeBERT has several advantages as compared with other code embedding techniques: First, different from traditional embedding techniques which produce context-independent embeddings, GraphCodeBERT generates contextual embeddings based on local code implementations. 
Moreover, instead of regarding the code snippet as a sequence of tokens, GraphCodeBERT uses data flow in pre-training stage, which captures the semantic-level code structure and encodes the intricate relationship between variables. 
GraphCodeBERT has been proven to be effective in many code-related tasks~\cite{guo2020graphcodebert, wan2022they}. 
}

\major{
Because our model inputs are triplets, {\sc TDPatcher} adopts three encoders based on GraphCodeBERT, i.e., \textit{Anchor Code-block Encoder}, \textit{Positive Code-block Encoder} and \textit{Negative Code-block Encoder}. 
These three encoders share weights of a single GraphCodeBERT as shown in Fig.~\ref{fig:workflow}. 
The GraphCodeBERT encoder contains 12 layers of transformers. 
The hidden size of each layer is 768. 
Each layer has a self-attention sub-layer which is composed of 12 attention heads. 
After feeding the code block sequence into the encoder, we extract the final hidden states $\mathbf{h}$ of the special token [CLS] as the code block embedding vector.
As a result, the three encoders map triplet inputs, i.e., anchor code block $\mathbf{A}$, positive code block $\mathbf{P}$, negative code block $\mathbf{N}$, into their corresponding code embeddings $\mathbf{h_{A}}$, $\mathbf{h_{P}}$ and $\mathbf{h_{N}}$ respectively.
}

\subsubsection{Contrastive Learning}
Now we present how to train {\sc TDPatcher} to distinguish equivalent code implementations from irrelevant implementations. 
The goal of our contrastive learning is that the positive pairs (e.g., equivalent code blocks) should be as close as possible in higher dimensional vector space, and the negative pairs should be as far away as possible in the space. 
% So far, the anchor code block $\mathbf{A}$, the positive code block $\mathbf{P}$, and the negative code block $\mathbf{N}$ are represented as independent contextual vectors, namely $\mathbf{h_A}$, $\mathbf{h_P}$, and $\mathbf{h_N}$ respectively. 
\major{
In this step, we introduce a triplet network to further learn the discriminative features among different inputs. 
The triplet network is a contrastive loss function based on the cosine distance operator~\cite{dong2018triplet, wei2022clear, hoffer2015deep}. 
}
The purpose of this function is to minimize the distances between the embeddings of similar code blocks (i.e., $\mathbf{h_A}$ and $\mathbf{h_P}$) and maximize the distances between the embeddings of irrelevant code blocks (i.e., $\mathbf{h_A}$ and $\mathbf{h_N}$). 
Formally, the training objective is to minimize the following loss function: 

\begin{equation}
\label{eq:patch}
max (||\mathbf{h_A} - \mathbf{h_P}|| - ||\mathbf{h_A} - \mathbf{h_N}|| + \epsilon, 0) 
\end{equation}
\major{where} % $\mathbf{h_a}$, $\mathbf{h_p}$ and $\mathbf{h_n}$ are the semantic embedings of the \textit{anchor} code block $\mathbf{A}$, the positive code block $\mathbf{P}$ and the negative code block $\mathbf{N}$. 
$\epsilon$ is the margin of the distance between $\mathbf{A}$ and $\mathbf{N}$. 
% By default, $\epsilon$ is set to 1, which means that the cosine distance between two irrelevant code blocks should be 1. 
Intuitively, the triplet loss encourages the cosine similarity between two equivalent code implementations to go up, and the cosine similarity between two irrelevant code blocks to go down.

\subsubsection{\major{TODO-missed} Methods Detection and Patching}
Based on the code embedding and contrastive learning techniques, we are able to apply our approach to solve the \textit{\major{TODO-missed} methods} detection and patching tasks. 
\begin{itemize}
    \item \major{Regarding the \textit{\major{TODO-missed} methods} detection task, for a codebase, we first extract all possible code blocks for each method, in this study, every consecutive five code lines is regarded as a code block (see the code block definition in Section~\ref{subsec:model_inputs}). 
    After that, we encode each code block into a vector by using our {\sc TDPatcher}. 
    When a new TODO comment is introduced, {\sc TDPatcher} first computes the vector representation for the suboptimal code block regarding the TODO comment, we then search code blocks in our codebase that are ``similar'' to the suboptimal code fragment. 
    Particularly, given vector representations of a code block and the suboptimal code fragment, if their cosine similarity score is over a pre-defined threshold, we determine the code block as a suboptimal code block. 
    The methods containing suboptimal code block(s) will be identified as \textit{\major{TODO-missed} methods}. 
    We set the pre-defined threshold to 0.9 in this study. 
    The threshold is a key parameter that governs whether a method will be predicted as a \textit{TODO-missed method}. 
    Grid search is employed to select the optimal threshold between 0.6 and 1.0 with a step size of 0.01.  
    We carefully tune the threshold on our validation set, the threshold to the best performance (i.e., F1-score) is selected as the optimal threshold (0.9 in this study) and used to evaluate our testing set. 
    } 
    \item \major{Regarding the \textit{\major{TODO-missed} methods} patching task, we aim to find the exact patching position for the TODO comment. 
    After identifying the \textit{TODO-missed method}, we divide the method into a list of code blocks (every consecutive five code lines) sequentially and choose the code block with the highest similarity score as the patching code block. 
    The central code line of the patching code block (i.e., the third code line within each code block) is regarded as the patching position for the \textit{TODO-missed method}. 
}
\end{itemize}

% \major{
% \subsection{Evaluation Setup}
% \label{sec:eval_setup}
% In this study, we constructed 10,775 triplet samples from 1,793 GitHub projects for training and evaluation. 
% We split the constructed triplet samples into three chunks by projects: 80\% projects' (i.e., 1,443 projects) triplet samples are used for training, 10\% (i.e., 175 projects) are used for validation and the rest (i.e., 175 projects) are held-out for testing. 
% As a result, our final training, validation and test sets consist of 8,531, 1,114, 1,130 triplet samples respectively. 
% It is worth mentioning that we split our dataset \textit{project-wise} instead of \textit{time-wise}. 
% In other words, we performed the cross-project evaluation settings~\cite{zimmermann2009cross, nam2013transfer, yan2018automating, qiu2021deep} for this study, which ensures our training samples, validation samples, and testing samples come from different projects. 
% The training set was used to fit the model for our task, while the validation set was used to find and optimize the best model under different hyperparameters. 
% The testing set was used for testing the final solution to confirm the actual predictive power of our model with optimal parameters.  
% }

% \section{Dataset Preparation}
% \label{sec:data}
% \input{data}

\section{Evaluations}
\label{sec:eval}
% https://github.com/CenterForOpenScience/osf.io/commit/6f6441b182dcb57dd4dc82cafa0861179a5ad869
% https://github.com/turicas/rows/commit/2a8e1c22c1b12ea7c1d272570b7fa46ef2fec942
% https://github.com/ckan/ckan/commit/b3381c13b935ec3cf84da28cec4f2f85378a1f52
% https://github.com/johnnykv/heralding/commit/4271d5b9b70d4a08ccc3c7322d814c0e302102e7
% https://github.com/datastax/python-driver/commit/e75dfe5ef746323b99050fe30d19a267badfbd4a
% https://github.com/sqlmapproject/sqlmap/blob/c5d20b8a86ad0a4718ef0c71364ebf59b732a9da/plugins/generic/takeover.py
% https://github.com/ckan/ckan/commit/b3381c13b935ec3cf84da28cec4f2f85378a1f52#
% https://github.com/uclnlp/jack/commit/916f4fad133d61a3efd4983a3786c96924dfe565
% We first present the baselines, the evaluation metrics and our experiment settings. 
% We then describe the results of our quantitative evaluation and manual analysis.
% Then, we describe , research questions (RQs), and the corresponding experimental results.
% \subsection{Research Questions}

In this section, we evaluate how well our approach for the task of \textit{\major{TODO-missed} methods} detection and patching.   
In particular, we first introduce the experimental setup for our work, we then aim to answer the following three key research questions:
\begin{itemize}
    \item \textbf{RQ-1:} How effective is {\sc TDPatcher} for \textit{\major{TODO-missed} methods} detection? 
    \item \textbf{RQ-2:} How effective is {\sc TDPatcher} for \textit{\major{TODO-missed} methods} patching?
    \item \textbf{RQ-3:} \major{How effective is {\sc TDPatcher} for using GraphCodeBERT and contrastive learning techniques?} 
    \item \major{\textbf{RQ-4:} How effective is {\sc TDPatcher} under different code block configurations?}
\end{itemize}

\subsection{Evaluation Setup}
\label{sec:eval_setup}
In this study, we constructed 10,775 triplet samples from 1,793 GitHub projects for training and evaluation. 
We split the constructed triplet samples into three chunks by projects: 80\% projects' (i.e., 1,443 projects) triplet samples are used for training, 10\% (i.e., 175 projects) are used for validation and the rest (i.e., 175 projects) are held-out for testing. 
As a result, our final training, validation and test sets consist of 8,531, 1,114, 1,130 triplet samples respectively. 
It is worth mentioning that we split our dataset \textit{project-wise} instead of \textit{time-wise}. 
In other words, we performed the cross-project evaluation settings~\cite{zimmermann2009cross, nam2013transfer, yan2018automating, qiu2021deep} for this study, which ensures our training samples, validation samples, and testing samples come from different projects. 
The training set was used to fit the model for our task, while the validation set was used to find and optimize the best model under different hyperparameters. 
The testing set was used for testing the final solution to confirm the actual predictive power of our model with optimal parameters.

\subsection{RQ-1. How effective is {\sc TDPatcher} for \major{TODO-missed} methods detection?}
\subsubsection{Experimental Setup.}
In this research question, we want to determine the effectiveness of our approach for identifying the \textit{\major{TODO-missed} methods}. 
The testing set contains 1,130 triplet samples. 
Each triplet relates to an anchor method $\mathbf{m_a}$, a positive method $\mathbf{m_p}$ and a negative method $\mathbf{m_n}$. 
One way to evaluate our approach is to see whether it can correctly predict $\langle \mathbf{m_a}, \mathbf{m_p} \rangle$ as positive and $\langle \mathbf{m_a}, \mathbf{m_n} \rangle$ as negative. 
In order to make the testing environment similar to the real-world working environment, we construct a candidate method pool $\mathcal{P}$ for evaluation. 
Theoretically, $\mathcal{P}$ should involve all methods within this project.
However, it is too expensive to do so due to the huge number of methods for large software projects. 
In our preliminary study, $\mathcal{P}$ is constructed by gathering all methods associated with the TODO-introduced files (more than 70\% \textit{TODO-introduced methods} are added within the same file).
In other words, the candidate method pool $\mathcal{P}$ for testing anchor method $\mathbf{m_a}$ contains $N$ methods, one of which is $\mathbf{m_p}$, where $N$ is the total number of extracted methods. 
For evaluation, we pair $\mathbf{m_a}$ with each candidate method in $\mathcal{P}$ and feed them into {\sc TDPatcher} to calculate a matching score, if the matching score is over a pre-defined threshold, the candidate method is regarded as a \textit{\major{TODO-missed} method.}

\subsubsection{Evaluation Metrics and Baselines.}
The \textit{\major{TODO-missed} methods} detection is a binary classification problem, we thus adopt the \textbf{Precision}, \textbf{Recall} and \textbf{F1-score} to measure the performance of our approach. 
To demonstrate the effectiveness of our proposed model, {\sc TDPatcher}, we compared it with the following chosen baselines:
\begin{itemize}
    \item \textbf{RG:} Random-Guess is usually selected as a baseline when the previous work is lacking~\cite{yan2018automating}. 
    For each candidate method, we randomly determine whether it is a \textit{\major{TODO-missed} method} or not. 
     
    \item \textbf{CEM:} Centrepiece-Exact-Match is a heuristic baseline to identify \textit{\major{TODO-missed} method.} 
    Since the TODO centrepiece is closely related to the TODO comment, we compare the TODO centrepiece with each line of the candidate method, if any lines match the TODO centrepiece exactly, then we declare this candidate method as a \textit{\major{TODO-missed} method.}
    
    \item \textbf{CSM:} Centrepiece-Soft-Match (\textbf{CSM}) modifies \textbf{CEM}
    by looking at the overlapping words between the TODO-centrepiece and the target line. 
    If the text similarity (proportion of the overlapped words) is above a threshold, we claim the candidate method as a     
    \textit{\major{TODO-missed} method.} 

    \item $\mathbf{NiCad:}$ NiCad~\cite{roy2008nicad} is a widely used traditional clone detection tool that handles a range of programming languages (e.g., C, Java, Python, C\#, PHP, Ruby, Swift) in terms of different granularities (i.e., method-level and block-level). 
    Given two methods are code clones, these two methods are likely to introduce TODO comments at the same time. 
    It is thus reasonable to adopt a clone detection method as a baseline. 
    For a given a \textit{TODO-introduced method}, the clone method will be regarded as its \textit{\major{TODO-missed} method.}
    In this study, we thus adopt NiCad as a tool to detect code clones.  % on both method-level and block-level, denoted as $\mathbf{NiCad_{M}}$ and $\mathbf{NiCad_{B}}$ respectively. 

    \item \major{\textbf{NCS:} Another thread of similar research that is relevant to our work is code search. 
    Identifying the \textit{TODO-missed method} can be viewed as searching code snippets related to \textit{TODO-introduced method}. 
    A plethora of approaches have been investigated for code search in software repositories~\cite{gu2018deep, zhang2019novel, sachdev2018retrieval, gao2023know}, and the key idea of these studies is mapping code to vectors and matching data in high dimensional vector space. 
    In this study, we choose NCS (Neural Code Search) model proposed by Facebook~\cite{sachdev2018retrieval} as the code search method baseline. 
    NCS encoded the code to vectors by combining FastText embeddings and TF-IDF weightings derived from the code corpus. 
    We can easily search for code ``similar'' to the \textit{TODO-introduced} method in their vector space. 
    } 
    
    \item \major{\textbf{Large Pre-trained Models:}
    Recently, the large pre-trained models, such as the CodeBERT~\cite{feng2020codebert}, PLBART~\cite{ahmad2021unified}, and CodeT5~\cite{wang-etal-2021-codet5}, have proven to be effective for many downstream tasks, including code generation, code summarization, code translation, code clone detection~\cite{tao2022c4, lu2021codexglue, kanade2020learning, ahmad2021unified, su2023still, yang2024federated}. 
    In this study, two large pre-trained models, i.e., CodeBERT and PLBART are chosen for our study. 
    We use CodeBERT and PLBART to encode code and detect code clones on method-level, denoted as $\mathbf{CodeBERT_{M}}$ and $\mathbf{PLBART_{M}}$ respectively. 
    Since {\sc TDPatcher} detects the \textit{\major{TODO-missed} methods} at the block-level (i.e., every consecutive five code lines of a method is regarded as a code block). 
    For a fairer comparison, we also use CodeBERT and PLBART to detect code clones at the same block-level, denoted as $\mathbf{CodeBERT_{B}}$ and $\mathbf{PLBART_{B}}$. 
    If any code blocks of a method are identified as code clones of a \textit{TODO-introduced method}, we claim this method as a \textit{\major{TODO-missed} method}. 
    }

    % \item \major{\textbf{PLBART:} 
    % PLBART is pre-trained on an extensive collection of Java and Python functions and associated NL text via denoising autoencoding.
    % }
    % \item $\mathbf{CodeBERT_{M}}:$ 
    % Recently, the large pre-trained based models, such as the CodeBERT model~\cite{feng2020codebert} has proven to be effective for many downstream tasks~\cite{tao2022c4, lu2021codexglue, kanade2020learning, ahmad2021unified}, including code clone detection. 
    % In this study, we use CodeBERT to encode code and detect code clones on method-level, denoted as $\mathbf{CodeBERT_{M}}$.
    % \item $\mathbf{CodeBERT_{B}}:$ {\sc TDPatcher} detects the \textit{\major{TODO-missed} methods} at the block-level (i.e., every consecutive five code lines of a method is regarded as a code block). 
    % For a fairer comparison, we use CodeBERT to detect code clones at same code block-level, denoted as $\mathbf{CodeBERT_{B}}$. 
    % If any code blocks of a method are identified as code clones of a \textit{TODO-introduced method}, we claim this method as a \textit{\major{TODO-missed} method}. 

\end{itemize}

\begin{table}%[tbp]
\caption{\major{TODO-missed} Methods Detection Evaluation}
\label{tab:rq1_eval_python}
% \vspace*{-10pt}
\major{
\begin{center}
\begin{tabular}{|l|c|c|c|c|}
    \hline
    {\bf Approach} & {\bf Precision} & {\bf Recall} & {\bf F1} \\
    \hline\hline
    \textbf{RG}  & $6.6\%$ & $49.7\%$ & $11.5\%$ \\
    \hline
    \textbf{CEM}  & $65.2\%$ & $39.2\%$ & $49.0\%$ \\
    \hline
    \textbf{CSM}  & $60.5\%$ & $46.1\%$ & $52.3\%$ \\
    \hline
    \textbf{NiCad} & $71.5\%$ & $35.4\%$ & $47.3\%$ \\
    \hline
    \textbf{NCS}  & $55.3\%$ & $54.2\%$ & $54.8\%$ \\
    \hline
    % $\mathbf{NiCad_B}$ & $73.3\%$ & $36.6\%$ & $49.3\%$ \\
    % \hline
    $\mathbf{PLBART_M}$ & $49.3\%$ & $47.6\%$ & $48.4\%$ \\
    \hline
    $\mathbf{CodeBERT_M}$ & $36.8\%$ & $56.5\%$ & $44.6\%$ \\
    \hline
    $\mathbf{PLBART_B}$  & $70.2\%$ & $68.8\%$ & $69.5\%$ \\
    \hline
    $\mathbf{CodeBERT_B}$ & $73.1\%$ & $68.3\%$ & $70.6\%$ \\
    \hline
    {\sc \textbf{TDPatcher}}  & $\textbf{83.7\%}$ & $\mathbf{77.3\%}$ & $\mathbf{80.4\%}$ \\
    \hline
\end{tabular}
% \vspace{-0.5cm}
\end{center}
}
\end{table}

\subsubsection{Experimental Results.}
The experimental results of our {\sc TDPatcher} and baselines are summarized in Table~\ref{tab:rq1_eval_python}. 
For each method mentioned above, the involved parameters were carefully tuned and the parameters with the best performance are used to report the final comparison results. 
From this table, we can observe the following points: 
RG achieves the worst performance regarding Precision, this is because our candidate method pool $\mathcal{P}$ has more than 16 methods on average, which shows the data imbalance problem in real scenarios.  
In general, the methods based on textual similarity, i.e., \textbf{CEM} and \textbf{CSM}, can achieve a relatively high precision score but low recall and F1 score. 
\major{
It is a little surprising that \textbf{CEM} achieves 65.2\% precision by naively checking the exact match with the TODO centrepiece statement, which indicates the TODO comments are closely related to the TODO centrepiece statements. 
}
\textbf{CSM} performs better than \textbf{CEM} in terms of F1-score and Recall, this is because \textbf{CSM} modifies the exact matching strategy with the soft matching by allowing a small proportion of different tokens, which also causes a performance drop regarding the Precision score. 
However, both \textbf{CEM} and \textbf{CSM} are based on bag-of-words matching, which can only capture the lexical level features, while the suboptimal implementations of \textit{\major{TODO-missed} methods} may be semantically related but lexically independent, which also explains their overall poor performance on Recall and F1 score.

% The traditional clone-detection tool \textbf{NiCad} achieves a relatively high precision but low recall, which explains its overall unsatisfactory performance. 
% This demonstrates that even the clone detection tool (i.e., \textbf{NiCad}) can accurately identify \textit{\major{TODO-missed} methods}, but it misses a large number of actual positive instances. 
% Because \textbf{NiCad} also shares the disadvantage of \textbf{CEM}/\textbf{CSM}, it detects code clones based on text similarities, which is unable to capture semantics of TODO comments and suboptimal code patterns. 
% Moreover, both method-level and block-level (e.g., for/while/if-statement blocks) of \textbf{NiCad} are too coarse to capture the critical context of the suboptimal implementations. 
\major{The traditional clone-detection tool \textbf{NiCad} and code search tool \textbf{NCS} achieve a relatively high precision but low recall, which explains its overall unsatisfactory performance. 
Both \textbf{NiCad} and \textbf{NCS} can be viewed as variants of the problem of finding ``similar'' code. 
In other words, \textbf{NiCad} and \textbf{NCS} search for code in a code base ``similar'' to a given piece of code (i.e., \textit{TODO-missed method}). 
The evaluation results show that even \textbf{NiCad} and \textbf{NCS} can accurately identify \textit{\major{TODO-missed} methods}, but they miss a large number of actual positive instances. 
\textbf{NCS} outperforms \textbf{NiCad} in terms of all evaluation metrics, this is because \textbf{NiCad} also shares the disadvantage of \textbf{CEM}/\textbf{CSM}, it detects code clones based on text similarities, which is unable to capture the semantics of TODO comments. 
Compared with \textbf{NiCad}, \textbf{NCS} encodes the code snippets into vector representations by using traditional word embedding techniques (i.e., FastText), which verifies the effectiveness of the idea of using code embeddings for this study. 
}

\major{
It is obvious that the large pre-trained models perform better on block-level (i.e., $\mathbf{CodeBERT_{B}}$ and $\mathbf{PLBART_{B}}$) than method-level (i.e., $\mathbf{CodeBERT_{M}}$ and $\mathbf{PLBART_{M}}$) by a large margin. 
Method-level approaches achieve a poor performance regarding the Precision and F1-score, the possible reason may be that the TODO comments are often associated with local suboptimal implementations, while the large pre-trained models encode the complete methods into vector representations, which brings a higher level of noise due to the very large size of method bodies. 
The block-level pre-trained models, i.e., $\mathbf{PLBART_{B}}$ and $\mathbf{CodeBERT_{B}}$, have their advantages as compared to the textual clone detection based method (i.e., \textbf{NiCad}) and code search based method (i.e., \textbf{NCS}). 
We attribute this to the following reasons: 
(i) First, compared with \textbf{NiCad} which treats words as discrete symbols and ignores the words' semantics, the code semantic features can be automatically encoded into numerical vectors by pre-trained models. 
(ii) Second, compared with \textbf{NCS} which uses traditional word embedding techniques, the large pre-trained models generate contextual embeddings. 
Specifically, the traditional word embedding techniques (e.g., word2vec, FastText) are context-independent, while the embeddings of pre-trained models are context-dependent. 
For example, regarding \textbf{NCS} model, for a given \texttt{print} statement (e.g., \texttt{print(`ip:'+internet.address)}), the same embeddings will be used for all instances of the same code pattern. 
However, even the commonly used \texttt{print} statement can serve as suboptimal implementations (e.g., \textit{TODO: remove this debug statement before releasing}) depending on its context. 
In contrast, CodeBERT and PLBART generate different embeddings for the commonly used \texttt{print} statement based on its code context. 
}

\major{
Our proposed model, {\sc \textbf{TDPatcher}}, outperforms all the baseline methods by a large margin in terms of all evaluation metrics. 
We attribute the superiority of our model to the following reasons: (i) We employ GraphCodeBERT to encode the inputs (including TODO comment and local code implementations) into contextual vectors. 
Similar to CodeBERT and PLBART, GraphCodeBERT takes advantage of large pre-trained models to generate contextual embeddings. 
(ii) GraphCodeBERT has its own advantage as compared to other large pre-trained models (e.g., CodeBERT and PLBART) by leveraging data flow graphs (DFG) to learn code representations. 
The use of data flow in GraphCodeBERT brings the following benefits to our task: a) data flow represents the dependency relation between variables, in other words, data flow represents the relation of ``where-the-value-comes-from'' between variables. 
This code structure provides crucial code semantic information for code understanding~\cite{guo2020graphcodebert}, especially in the context of our local code implementations. 
For example, as illustrated in \texttt{\_check\_files()} in  Fig.~\ref{fig:moti_example2}, linking the suboptimal variable \texttt{line} with \texttt{subprocess} directly is challenging, data flow provides a better way to understand the variable \texttt{line} comes from \texttt{proc}, while the value of \texttt{proc} comes from \texttt{subprocess}. 
b) Compared with AST, data flow is less complex and does not bring an unnecessarily deep hierarchy, which can bring performance boosts on varying sequence lengths. 
(iii) Moreover, by incorporating the contrastive learning strategy, {\sc \textbf{TDPatcher}} can better align equivalent suboptimal implementations across different code blocks. 
The semantic and discriminative features learned from the triplet inputs assist {\sc TDPatcher} in detecting \textit{TODO-missed methods}. 
The aforementioned techniques collectively contribute to the remarkable performance of our proposed model.
}

\find{
{\bf Answer to RQ-1: How effective is {\sc TDPatcher} for \textit{\major{TODO-missed} methods} detection?}
We conclude that our approach is highly effective for the task of detecting \textit{\major{TODO-missed} methods} in software projects.
}

% Considering the \textbf{CEM} and \textbf{CSM} determine the \textit{TODO-dropped methods} naively based on exact/partial match with respect to the TODO centrepiece. the 

\subsection{RQ-2. How effective is {\sc TDPatcher} for \major{TODO-missed} methods patching?}
\subsubsection{Experimental Setup.}
In this research question, we want to evaluate the effectiveness of our approach for patching the \textit{\major{TODO-missed} methods.}
In other words, if a \textit{\major{TODO-missed} method} needs to be updated (i.e., adding the TODO comment), whether {\sc TDPatcher} can patch the TODO comment to the right position. 
\major{
In particular, for each testing triplet sample $\langle \mathbf{m_a}, \mathbf{m_p}, \mathbf{m_n} \rangle$, since the anchor method $\mathbf{m_a}$ and the positive method $\mathbf{m_p}$ contain the same TODO comment, we take the anchor method $\mathbf{m_a}$ as the \textit{TODO-introduced method}, and the positive method $\mathbf{m_p}$ as our ground truth (i.e., \textit{TODO-introduced method}).  
Given the \textit{TODO-introduced method} $\mathbf{m_a}$, we evaluate our approach by checking how often the expected patching position of the \textit{TODO-missed method} $\mathbf{m_p}$ can be accurately found and recommended. 
To be more specific, we first divide the positive method $\mathbf{m_p}$ into a list of code blocks sequentially.  
We then pair the anchor code block of $\mathbf{m_a}$ with each divided code block of $\mathbf{m_p}$ and estimate a similarity score by applying our model. 
The divided code blocks are ranked by their similarity scores for recommendation. 
}
\major{
If the expected patching position lies within the recommended code block, we consider {\sc TDPatcher} patches the \textit{TODO-missed method} $\mathbf{m_p}$ successfully. 
Notably, we ignore the $\pm2$ lines differences for the task of \textit{TODO-missed method} patching, because we easily determine the exact TODO patching position once the target code block is accurately suggested. 
}

\subsubsection{Evaluation Metrics and Baselines.}
\major{
The \textit{\major{TODO-missed} method} patching task is a ranking problem, we thus adopted the widely-accepted metric, $\mathbf{P@K}$~\cite{liu2009learning} and $\mathbf{DCG@K}$~\cite{jarvelin2002cumulated} to measure the ranking performance of our model. 
$\mathbf{P@K}$ is the precision of the expected patching position in top-K recommended locations. 
Different from $\mathbf{P@K}$, $\mathbf{DCG@K}$ gives a higher reward for ranking the correct element at a higher position. 
}
\major{
To demonstrate the effectiveness of our {\sc TDPatcher} for \textit{\major{TODO-missed} method} patching task, we compare it with the $\mathbf{CodeBERT_B}$ baseline (denoted as  $\mathbf{CB_B}$ for short in Table~\ref{tab:rq2_eval_python}) and \textbf{NCS} baseline due to their superiority among large pre-trained models and similarity matching based models in RQ-1 respectively. 
% its ability to locate the patching position for a \textit{TODO-missed} method.  
}

\subsubsection{Experimental Results.}
\major{
The evaluation results of our approach {\sc TDPatcher} and baseliens (including \textbf{NCS} and $\mathbf{CodeBERT_B}$) for \textbf{RQ-2} are summarized in Table~\ref{tab:rq2_eval_python}. 
From the table, we can see that: 
(i) \textbf{NCS} achieves the worst performance regarding $\mathbf{P@K}$ and $\mathbf{DCG@K}$. 
For similarity matching based approaches, they heavily rely on whether similar code snippets can be found and how similar the code snippets are. 
Since equivalent suboptimal implementations may have different contexts, it is difficult to identify the corresponding suboptimal locations by simply measuring text similarities. 
}
\major{(ii)} {\sc TDPatcher} can patch the TODO comment to the \textit{\major{TODO-missed} method} effectively. 
For example, for a given \textit{\major{TODO-missed} method}, {\sc TDPatcher} can successfully retrieve the right patching position in the first place with a probability of \major{71\%}, this score increases up to \major{92.5\%} when we enlarge the number of recommendations from 1 to 5, which shows the advantage of our approach for identifying the suboptimal code implementations. 
\major{(iii)} As can be seen, our model {\sc TDPatcher} is stably and substantially better than the $\mathbf{CodeBERT_B}$ baseline in terms of $\mathbf{P@K}$ and $\mathbf{DCG@K}$. 
Both the baseline method, i.e., $\mathbf{CodeBERT_B}$, and our approach can be viewed as variants of embedding algorithm(s), which map code blocks into vectors of a high-dimensional vector space and calculate the similarity scores between vectors. 
Therefore, the key of \textit{\major{TODO-missed} methods} patching relies on how good the generated embeddings are for learning the implicit semantic features.  
{\sc TDPatcher} has its advantage over $\mathbf{CodeBERT_B}$ with respect to the following two aspects: First, the embeddings generated by GraphCodeBERT are more suitable than CodeBERT by incorporating the data-flow to capture the inherent structure of code; % for semantic mapping between code and natural languages; 
Moreover, the contrastive learning further adjusts the embeddings for learning equivalent code implementations. 
The superior performance also verifies the embeddings generated by our approach convey a lot of valuable information.

\find{
{\bf Answer to RQ-2: How effective is {\sc TDPatcher} for \textit{\major{TODO-missed} methods} patching?}
We conclude that our approach is effective for pinpointing the exact code lines where TODO comments need to be added, which verifies the effectiveness of our approach for the task of \textit{\major{TODO-missed} methods} patching.
}

\begin{table}%[tbp]
\caption{\major{TODO-missed} Methods Patching Evaluation}
\label{tab:rq2_eval_python}
% \vspace*{-10pt}
\begin{center}
\major{
\begin{tabular}{||c|c|c|c||c|c|c|c||}
    \hline
    {\bf Measure} & {\bf NCS} & {\bf $\mathbf{CB_B}$} & {\bf Ours} & {\bf Measure} & {\bf NCS} & {\bf $\mathbf{CB_B}$} & {\bf Ours} \\
    \hline\hline
    \textbf{P@1}  & $50.4\%$ & $60.8\%$ & $\mathbf{71.0\%}$ & \textbf{DCG@1} & $50.4\%$ & $60.8\%$ & $\mathbf{71.0\%}$ \\
    \hline
    \textbf{P@2}  & $60.0\%$ & $71.3\%$ & $\mathbf{81.1\%}$ & \textbf{DCG@2} & $56.4\%$ & $67.4\%$ & $\mathbf{77.3\%}$ \\
    \hline
    \textbf{P@3}  & $66.3\%$ & $76.5\%$ & $\mathbf{86.8\%}$ & \textbf{DCG@3} & $59.6\%$ & $70.0\%$ & $\mathbf{80.2\%}$ \\
    \hline
    \textbf{P@4}  & $70.5\%$ & $79.3\%$ & $\mathbf{90.3\%}$ & \textbf{DCG@4} & $61.4\%$ & $71.2\%$ & $\mathbf{81.7\%}$\\
    \hline
    \textbf{P@5}  & $74.0\%$ & $82.7\%$ & $\mathbf{92.5\%}$ & \textbf{DCG@5} & $62.8\%$ & $72.6\%$ & $\mathbf{82.6\%}$ \\
    \hline
    % {\sc \textbf{TDCleaner}}  & $\mathbf{85.0\%}$ & $\textbf{86.2\%}$ & $\mathbf{84.4\%}$ & $\mathbf{85.3\%}$ \\
    % \hline
\end{tabular}
% \vspace{-0.5cm}
}
\end{center}
\end{table}

\subsection{RQ-3. How effective is {\sc TDPatcher} for using GraphCodeBERT and contrastive learning?}

\subsubsection{Experimental Setup.}
To better capture the semantic correlations between code blocks and TODO comments, we employ the \textbf{GraphCodeBERT} as encoders and \textbf{contrastive learning} for acquiring suitable vector representations. 
In this research question, we conduct experiments to verify their effectiveness one by one. 

\subsubsection{Evaluation Metrics and Baselines.} 
To verify the effectiveness of using \textbf{GraphCodeBERT} and \textbf{contrastive learning}, we compare {\sc TDPatcher} with the following two baselines on \textit{\major{TODO-missed} method} detection (RQ-1) and patching (RQ-2) tasks respectively, we use the same evaluation metrics for RQ-1 and RQ-2. 
\begin{itemize}
    \item \textbf{CLO} (Contrastive Learning Only) removes GraphCodeBERT from our model and only retains \textbf{contrastive learning}. 
    We replace GraphCodeBERT embedding with traditional word embedding techniques (i.e., Glove word vectors). 
    The \textbf{CLO} is then trained with the triplet samples with contrastive learning. 
    \item \major{\textbf{GCBO} (GraphCodeBERT only) removes contrastive learning from our model and only keeps the \textbf{GraphCodeBERT} as encoders, the generated vectors of code blocks are used to calculate the similarity score directly without going through the contrastive learning layer. 
    \textbf{GCBO} is similar to $\mathbf{CodeBERT_B}$ baseline except its code embeddings are generated by \textbf{GraphCodeBERT} instead of \textbf{CodeBERT}. 
    }
\end{itemize}

\subsubsection{Experimental Results.} 
The experimental results of the ablation analysis on RQ-1 and RQ-2 are summarized in Table~\ref{tab:ab_eval_rq1} and Table~\ref{tab:ab_eval_rq2} respectively. 
It can be seen that: 
(i) No matter which component we removed, it hurts the overall performance of our model, which verifies the usefulness and necessity of using \textbf{GraphCodeBERT} and \textbf{contrastive learning.}
(ii) \textbf{CLO} achieves the worst performance. 
There is a significant drop overall in every evaluation metric after removing the GraphCodeBERT.
This signals that the embeddings generated by GraphCodeBERT have a major influence on the overall performance. 
(iii) \major{
The only difference between \textbf{GCBO} and $\mathbf{CodeBERT_B}$ is their code embedding generation model (e.g., \textbf{GCBO} uses GraphCodeBERT while $\mathbf{CodeBERT_B}$ uses CodeBERT). 
Even though \textbf{GCBO} doesn't get top results as {\sc TDPatcher}, it still achieves a considerable performance on RQ-1 and RQ-2 (better than $\mathbf{CodeBERT_B}$ baseline), which further confirms the advantage of GraphCodeBERT embeddings over CodeBERT.
}
(vi) By comparing the performance of \textbf{GCBO} with our {\sc TDPatcher}, we can measure the performance improvement achieved by employing the contrastive learning. 
It is notable that the advantage of our model is more obvious on RQ-2, which shows the benefits of contrastive learning for separating suboptimal code implementations from irrelevant ones.

\begin{table}%[tbp]
\caption{Ablation Evaluation on RQ-1}
\label{tab:ab_eval_rq1}
% \vspace*{-10pt}
\major{
\begin{center}
\begin{tabular}{|l||c|c|c|}
    \hline
    \textbf{Measure} & \textbf{CLO} & \textbf{GCBO} & {\sc \textbf{TDPatcher}} \\
    \hline\hline
    \textbf{Precision} & $55.0\%$  & $75.0\%$ & $\mathbf{83.7\%}$ \\
    \hline
    \textbf{Recall} & $65.3\%$ & $74.0\%$ & $\mathbf{77.3\%}$ \\
    \hline
    \textbf{F1} & $59.7\%$ & $74.4\%$ & $\mathbf{80.4\%}$ \\
    \hline
\end{tabular}
\end{center}
}
% \vspace{-6pt}
\end{table}

\begin{table}%[tbp]
\caption{\major{Ablation Evaluation on RQ-2}}
\label{tab:ab_eval_rq2}
% \vspace*{-10pt}
\begin{center}
\major{
\begin{tabular}{|c|c|c|c||c|c|c|c|}
    \hline
    \textbf{Measure} & \textbf{CLO} & \textbf{GCBO} & {\sc \textbf{TDPatcher}} & \textbf{Measure} & \textbf{CLO} & \textbf{GCBO} & {\sc \textbf{TDPatcher}} \\
    \hline\hline
    \textbf{P@1} & $56.2\%$ & $63.1\%$ & $\mathbf{71.0\%}$ & \textbf{DCG@1} & $56.2\%$ & $63.1\%$ & $\mathbf{71.0\%}$ \\
    \hline
    \textbf{P@2} & $65.5\%$ & $74.4\%$ & $\mathbf{81.1\%}$ & \textbf{DCG@2} & $62.0\%$ & $70.2\%$ & $\mathbf{77.3\%}$ \\
    \hline
    \textbf{P@3} & $68.8\%$ & $80.2\%$ & $\mathbf{86.8\%}$ & \textbf{DCG@3} & $63.7\%$ & $73.1\%$ & $\mathbf{80.2\%}$ \\
    \hline
    \textbf{P@4} & $74.1\%$ & $82.3\%$ & $\mathbf{90.3\%}$ & \textbf{DCG@4} & $66.0\%$ & $74.0\%$ & $\mathbf{81.7\%}$ \\
    \hline
    \textbf{P@5} & $76.9\%$ & $85.0\%$ & $\mathbf{92.5\%}$ & \textbf{DCG@5} & $67.1\%$ & $75.1\%$ & $\mathbf{82.6\%}$ \\
    \hline
\end{tabular}
}
\end{center}
% \vspace{-6pt}
\end{table}

\find{
{\bf Answer to RQ-3: How effective is {\sc TDPatcher} for for using GraphCodeBERT and contrastive learning?}
We conclude that both GraphCodeBERT and contrastive learning are effective and helpful to enhance the performance of our model. 
}

\subsection{\major{RQ-4. How effective is {\sc TDPatcher} under different code block configurations?}}
\subsubsection{Experimental Setup.}
\major{
In our study, {\sc TDPatcher} encodes the TODO comment and its associated suboptimal implementations by using code blocks. 
Each code block is constructed with three segments: namely \textit{TDOO comment}, \textit{TODO centrepiece}, \textit{TODO context}. 
Particularly, we choose the code line after the TODO comment as the \textit{TODO centrepiece}, and choose two code lines before and after TODO comment (excluding the \textit{TODO centrepiece}) as the \textit{TODO context}. 
% The \textit{TODO centrepiece} and \textit{TODO context} are chosen based on our empirical observations, there is no guarantee that the current code block settings are sound and optimal. 
Our choice of \textit{TODO centrepiece} and \textit{TODO context} is based on empirical observations. 
There is no assurance that the current configuration of code blocks is definitive and/or optimal. 
In this research question, we aim to investigate the {\sc TDPatcher}'s performance across various code block configurations. 
}

\subsubsection{Evaluation Metrics and Baselines.} 
\major{
We estimate {\sc TDPatcher}'s performance by adjusting code block configurations using different \textit{TODO centrepiece} and \textit{TODO context} settings. 
Specifically, to evaluate the impact of different \textit{TODO centrepiece} choices, we keep the \textit{TODO context} settings consistent with the previous configuration. 
We then compare the {\sc TDPatcher} against the following variants on \textit{TODO-missed method} detection (RQ-1) and patching tasks (RQ-2) respectively. 
The evaluation metrics employed for both RQ-1 and RQ-2 remain the same for a fair comparison. 
\begin{itemize} 
    \item \textbf{CEN-1:} For this baseline, we use the code line before the TODO comment as the \textit{TODO centrepiece}. 
    The \textit{TODO context} remains consistent with the settings used in TDPatcher. 
    This baseline is denoted as \textbf{CEN-1}. 
    \item \textbf{CEN+1:} For this baseline, we use the second code line after the TODO comment as the \textit{TODO centrepiece}. 
    The \textit{TODO context} remains the same settings with {\sc TDPatcher}. 
    This baseline is denoted as \textbf{CEN+1}. 
\end{itemize}
Similarly, to assess different configurations of \textit{TODO context}, we keep \textit{TODO centrepiece} the same as previous settings (i.e., the code line following the TODO comment) and compare {\sc TDPatcher} with the following \textit{TODO context} variants. 
\begin{itemize}
    \item \textbf{CON\_1}: We use one code line before and code one line after the TODO comment (excluding the \textit{TODO centrepiece}) as \textit{TODO context}, this baseline is denoted as \textbf{CON\_1}. 
    \item \textbf{CON\_3}: We use three code lines before and three code lines after the TODO comment (excluding the \textit{TODO centrepiece}) as \textit{TODO context}, this baseline is denoted as \textbf{CON\_3}. 
\end{itemize}
}

\begin{table}%[tbp]
\caption{\major{Different Code Block Evaluation on RQ-1}}
\label{tab:code_block_eval_rq1}
% \vspace*{-10pt}
\major{
\begin{center}
\begin{tabular}{|l|c|c|c|c|}
    \hline
    {\bf Approach} & {\bf Precision} & {\bf Recall} & {\bf F1} \\
    \hline\hline
    \textbf{CEN-1}  & $78.2\%$ & $74.1\%$ & $76.1\%$ \\
    \hline
    \textbf{CEN+1}  & $78.2\%$ & $73.6\%$ & $75.7\%$ \\
    \hline
    \textbf{CON\_1}  & $77.4\%$ & $70.6\%$ & $73.6\%$ \\
    \hline
    \textbf{CON\_3} & $71.4\%$ & $74.8\%$ & $73.1\%$ \\
    \hline
    {\sc \textbf{TDPatcher}}  & $\textbf{83.7\%}$ & $\mathbf{77.3\%}$ & $\mathbf{80.4\%}$ \\
    \hline
\end{tabular}
% \vspace{-0.5cm}
\end{center}
}
\end{table}

\begin{table}%[tbp]
\caption{\major{Different Code Block Evaluation on RQ-2}}
\label{tab:code_block_eval_rq2}
% \vspace*{-10pt}
\begin{center}
\major{
\begin{tabular}{|c|c|c|c|c|c|}
    \hline
    \textbf{Measure} & \textbf{CEN-1} & \textbf{CEN+1} & \textbf{CON\_1} & \textbf{CON\_3} & {\sc \textbf{TDPatcher}} \\
    \hline\hline
    \textbf{P@1} & $60.7\%$ & $64.5\%$ & $65.3\%$ & $64.0\%$ & $\mathbf{71.0\%}$ \\
    \hline
    \textbf{P@2} & $68.8\%$ & $76.8\%$ & $76.0\%$ & $75.1\%$ &  $\mathbf{81.1\%}$ \\
    \hline
    \textbf{P@3} & $74.5\%$ & $81.2\%$ & $81.1\%$ & $80.1\%$ & $\mathbf{86.8\%}$ \\
    \hline
    \textbf{P@4} & $78.2\%$ & $84.3\%$ & $84.5\%$ & $83.5\%$ & $\mathbf{90.3\%}$ \\
    \hline
    \textbf{P@5} & $80.7\%$ & $87.3\%$ & $88.2\%$ & $86.9\%$ & $\mathbf{92.5\%}$ \\
    \hline\hline
    \textbf{DCG@1} & $60.7\%$ & $64.5\%$ & $65.3\%$ & $64.0\%$ & $\mathbf{71.0\%}$ \\
    \hline
    \textbf{DCG@2} & $65.8\%$ & $72.2\%$ & $72.1\%$ & $71.0\%$ &  $\mathbf{77.3\%}$ \\
    \hline
    \textbf{DCG@3} & $68.7\%$ & $74.4\%$ & $74.6\%$ & $73.5\%$ & $\mathbf{80.2\%}$ \\
    \hline
    \textbf{DCG@4} & $70.3\%$ & $75.8\%$ & $76.1\%$ & $75.0\%$ & $\mathbf{81.7\%}$ \\
    \hline
    \textbf{DCG@5} & $71.2\%$ & $77.0\%$ & $77.5\%$ & $76.3\%$ & $\mathbf{82.6\%}$ \\
    \hline
\end{tabular}
}
\end{center}
% \vspace{-6pt}
\end{table}

\subsubsection{Experimental Results.} 
\major{The experimental results of {\sc TDPatcher} under different code block settings are shown in Table~\ref{tab:code_block_eval_rq1} and Table~\ref{tab:code_block_eval_rq2} for RQ1 and RQ2 respectively. 
From the table, several points stand out: 
(i) Regarding the \textit{TODO centerpiece}, we notice that our framework achieves its best performance when the \textit{TODO centerpiece} is set to the code line following the TODO comment. 
This is reasonable because putting comments before the code implementations is a common practice for software development.
\textbf{CEN+1} outperforms \textbf{CEN-1} for both RQ1 and RQ2 also confirms our assumption that the code lines after the TODO comment are more closely related to the suboptimal implementations. 
(ii) Regarding the \textit{TODO context}, {\sc TDPatcher} has its advantage as compared to \textbf{CON\_1} and \textbf{CON\_3}. 
\textbf{CON\_1} only includes one code line before and after the TODO comment as its TODO context, in other words, each code block of \textbf{CON\_1} only contains three code lines.  
The poor performance of \textbf{CON\_1} shows that it is not sufficient to cover the suboptimal implementations associated with the TODO comment. 
\textbf{CON\_3} contains three code lines before and after the TODO comment as its \textit{TODO context}. 
The overall performance decreases when we increase the code block from 5 code lines to 7 code lines. 
This suggests that larger code block settings bring more noise into the code block in the form of irrelevant code lines, and it thus incurs bigger challenges and difficulties for our task of \textit{TODO-missed Methods} detection and patching. 
}

\find{
{\bf \major{Answer to RQ-4: How effective is {\sc TDPatcher} under different code block configurations?}}
\major{
We conclude that {\sc TDPatcher} achieves its optimal performance when we set the \textit{TODO centerpiece} as the code line following the TODO comment, and the \textit{TODO context} as two code lines surrounding the \textit{TODO centerpiece}. 
}
}

% The key to the \textit{TODO-dropped methods} detection and patching is to capture the semantic and discriminative features between TODO comments and code blocks,  

% \subsection{Manual Analysis}
% https://github.com/openwisp/netjsonconfig/commit/7a152a344333665bbc9217011418ae39e8a1af81
% https://github.com/CenterForOpenScience/osf.io/commit/341ea7df3a4eee5a66bcbb440e7a6a28a599e9ef
% https://github.com/openstack/trove/commit/0ee399a82835c2a57e015058025bc094def0d92f
% https://github.com/uclnlp/jack/blob/43f4b30ba5bb337f8ba92b02cfce71efd1b939a8/jtr/jack/readers.py
% !!! https://github.com/explosion/spaCy/commit/187f37073495211c422be719b16da4d2449c8844
% https://github.com/CenterForOpenScience/osf.io/blob/108dee325f607fee0e07c05fdc1a02319a36e747/tests/test_views.py
% https://github.com/rapidsms/rapidsms/commit/a0a3d04369716cddc3d79c39136d7e892b897f0c
% !!! https://github.com/ckan/ckan/commit/29f6d03946cd12656b1b75b2e17d10873c9ccafc
% https://github.com/inducer/loopy/commit/82a6a4afb170dba9f645576224f449574ee347ac
% https://github.com/ClusterHQ/flocker/commit/e4c91fcbc60ec167a774e1250919340fd700abbc
% https://github.com/Yelp/osxcollector/commit/790eb8a695f930f4561718cf860ebf539352903a
% https://github.com/biocore/qiime/commit/05c77442cfc4617568fd72c19445907275a2afb9
% https://github.com/mathics/Mathics/commit/633ae18f00067c40420049cd415b30aec44db4c1

\subsection{Manual Analysis}
\begin{figure}
% \vspace{-7pt}
\centerline{\includegraphics[width=0.85\textwidth]{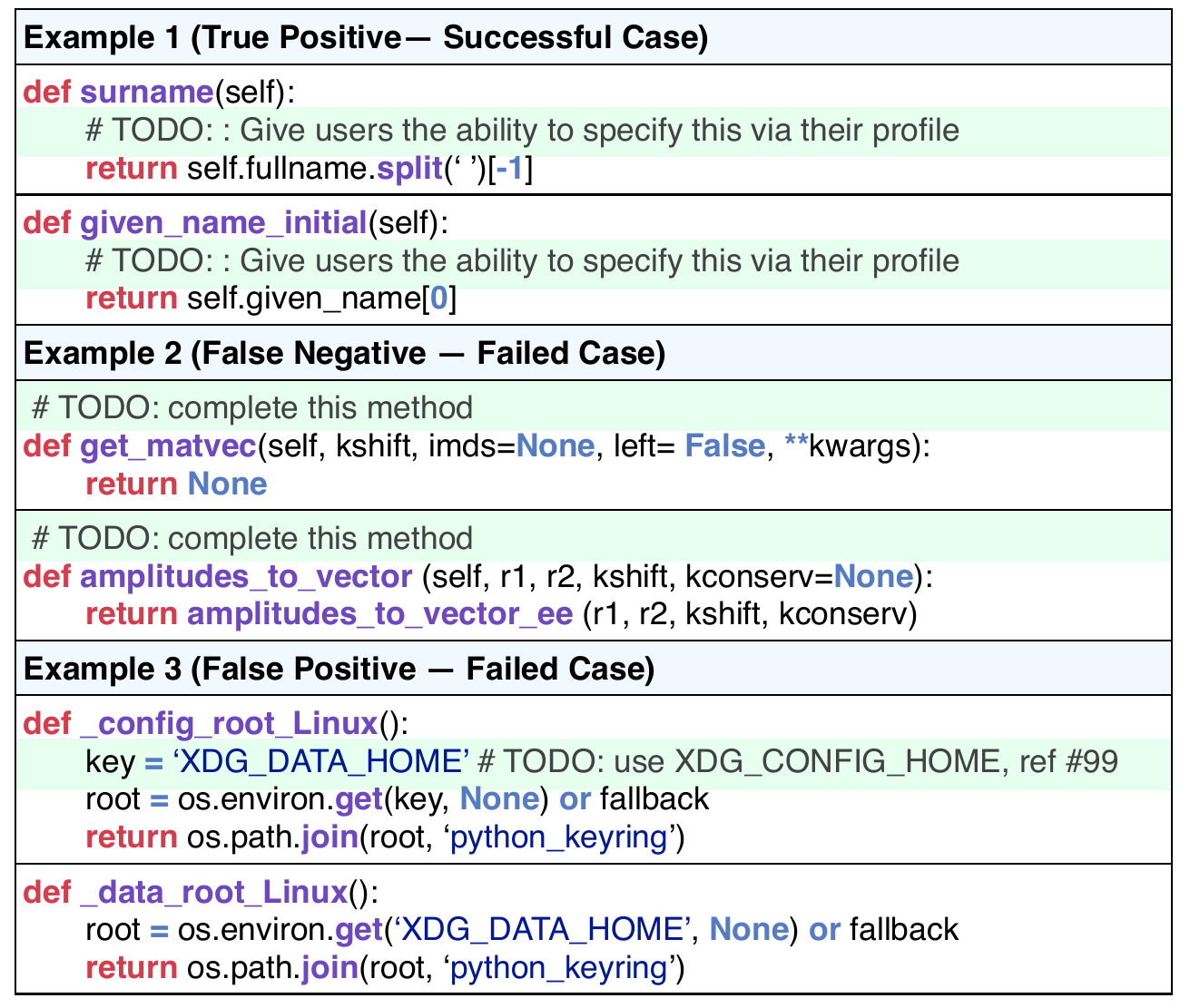}}
% \vspace*{-5pt}
\caption{Manual Analysis Examples}
\label{fig:manual_example}
% \vspace{-8pt}
\end{figure}

To better understand the strengths and limitations of our approach, we manually inspect a number of test results.
Example 1 in Fig.~\ref{fig:manual_example} shows an example where {\sc TDPatcher} makes correct predictions while other baselines fail.
From this example, we can observe that {\sc TDPatcher} successfully learns to capture the semantic-level features between TODO comments and code implementations, while the textual similarity based models rely on lexical-level features and are hard to model such correlations.
Moreover, contrastive learning further enables our {\sc TDPatcher} to learn implicit connections among similar code patterns.
% https://github.com/in-toto/in-toto/commit/be95285aaff3a494f05c666d857a17d4f779c78e
% https://github.com/openstack/trove/commit/0ee399a82835c2a57e015058025bc094def0d92f
% https://github.com/biocore/qiime/commit/05c77442cfc4617568fd72c19445907275a2afb9
% https://github.com/CenterForOpenScience/osf.io/commit/6f6441b182dcb57dd4dc82cafa0861179a5ad869
% !!! https://github.com/sqlmapproject/sqlmap/commit/a1b83cd56f231e70be931d270cbc65b2638110ed
% https://github.com/turicas/rows/commit/a2ae1a2ebeb823d750cdf2422fe6e07ae6af0b6a

% These samples presented 
% We summarize two common aspects that {\sc TDPatcher} is difficult to deal with. 
\major{
We summarize two aspects that cause {\sc TDPatcher} to experience difficulties}.
First of all, one common failed situation is that some suboptimal implementations are too general to be matched, it is very difficult, if not possible, for {\sc TDPatcher} to correctly infer such test samples. 
For example, as shown in Example 2 of Fig.~\ref{fig:manual_example}, the developer added a TODO comment ``\textit{TODO: complete this method}'' to different unfinished methods at the same time (i.e., \texttt{get\_matvec} and \texttt{amplitudes\_to\_vector}), which reminds the developer to work on these two methods when time permits. 
{\sc TDPatcher} failed to detect \texttt{amplitudes\_to\_vector} missing the above TODO comment, this is because it is not easy to claim this method is completed or not under the current code context. 
% due to his/her unfamiliarity with the code. 

Another bad situation is that the suboptimal implementations do not provide sufficient information for \textit{\major{TODO-missed} methods} identification.  
Example 3 in Fig.~\ref{fig:manual_example} shows such a test sample, the developers added a TODO comment, i.e., ``\textit{TODO: use XDG\_CONFIG\_HOME}'' to \texttt{\_config\_root\_Linux} method to indicate the temporary implementation for the \texttt{key} variable.   
However, this code implementation is suboptimal for the method  \texttt{\_config\_root\_Linux} while complete for the method \texttt{\_data\_root\_Linux}. 
Considering our approach lacks the contextual information regarding method names, {\sc TDPatcher} thus mispatched this TODO comment to \texttt{\_data\_root\_Linux} method. 
% {\sc TDPatcher} mispatched this TODO comment to \texttt{\_data\_root\_Linux} method considering the same code implementations of these two methods. 
% are too trivial and/or complicated for {\sc TDPatcher} to perfectly handle. 
% ``\textit{TODO: enable this check when \#1502 is fixed}'' to \texttt{test\_group\_read} method to indicate themselves to uncomment the following $\mathbf{assert}$ statement when the issue is fixed. 
% because our approach correlates the TODO comment with the assert statement while ignores the non-trivial differences, i.e., the $\mathbf{assert}$ statement is invalid (uncommented) in  \texttt{test\_group\_read} but valid in \texttt{test\_group\_index} method.
Addressing these challenges will remarkably boost the learning performance of our model, we will focus on this research direction in the future. 

\find{
{\bf Why does {\sc TDPatcher} succeed/fail?}
{\sc TDPatcher} successfully learns to capture the semantical features between TODO comments and their suboptimal implementations, however, it fails to handle if the suboptimal implementations are too general or lacking sufficient information, we will try to address these limitations in our future work. 
}

% \vspace{-8pt}

\section{In-the-wild Evaluation}
\label{sec:disc}
% https://github.com/turicas/rows/commit/a27f499d25a037a123358b67ca2e1d5f407d9ebe
% https://github.com/oduwsdl/ipwb/commit/7450e406684d7739e82b1634533af05f11fe0820
% https://github.com/poliastro/poliastro/commit/9e4033592a18ca899347221a4912badb222ce0d4
% https://github.com/chipmuenk/pyfda/commit/a2b877011af7f30e14c8d99fa32f1daa8a12aef8
% https://github.com/cms-dev/cms/commit/77610b3801de6b8d150fcd9ca3879df584e0dcc3
% https://github.com/bcbio/bcbio-nextgen/commit/73e9c9698ba2110bacd4081d1e7eee9698d734c2
% https://github.com/thunder-project/thunder/commit/7bffa41b3e29112f5c0abe45f8cc738ad418a9f7
% https://github.com/ansible/ansible/commit/8f43d222c04895809cbc6b2fee94e66950293719
% https://github.com/simpeg/simpeg/commit/581f1b15f137e414fc3ed628ffaf56c2f082ec23
% 

\begin{figure}
% \vspace{-7pt}
\centerline{\includegraphics[width=0.85\textwidth]{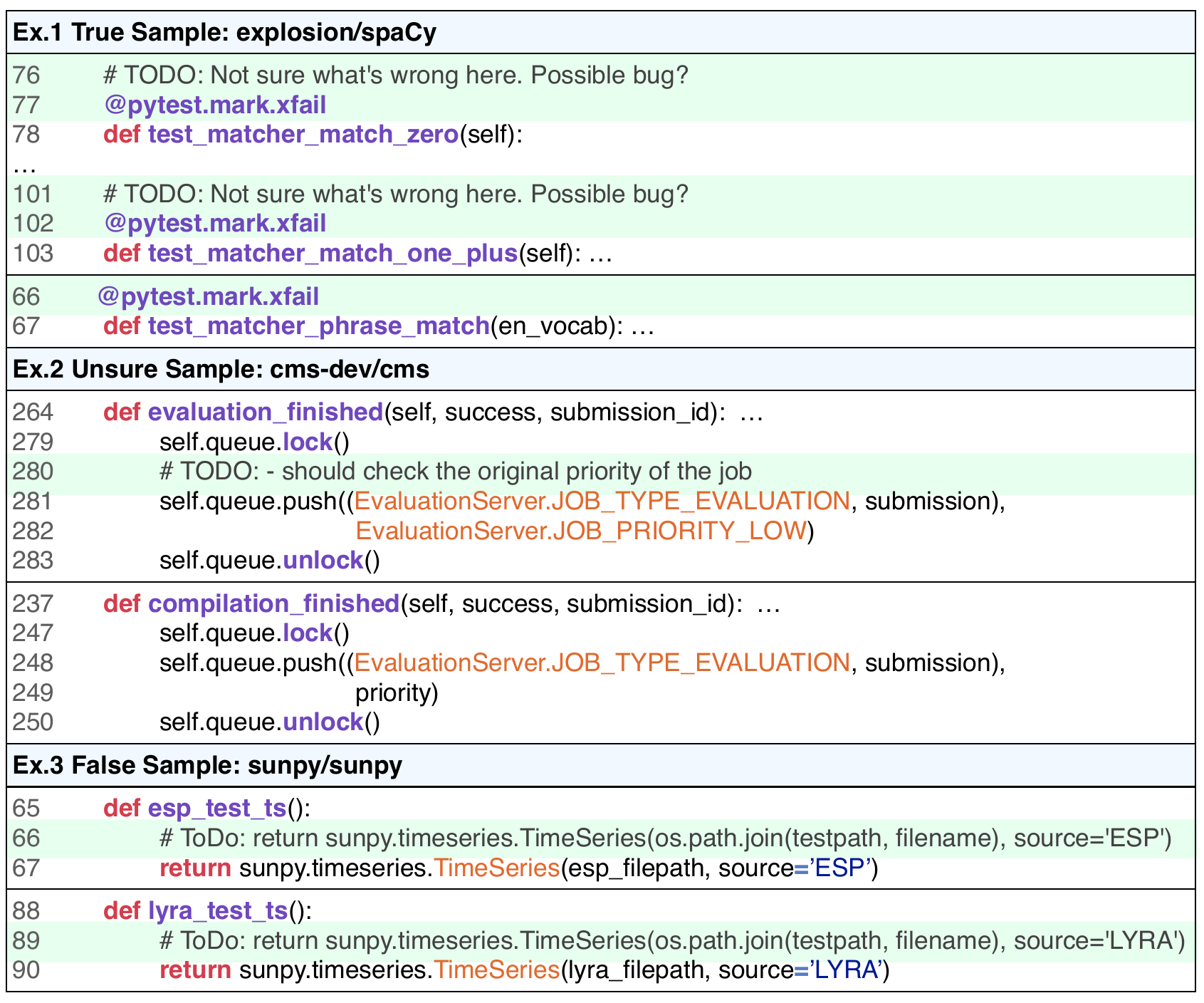}}
% \vspace*{-5pt}
\caption{In-the-wild Evaluation Examples}
\label{fig:in_the_wild_example}
% \vspace{-8pt}
\end{figure}

\major{Besides the automatic evaluation, we further conduct an in-the-wild evaluation to evaluate the effectiveness of {\sc TDPatcher} in real-world software projects. 
To do so, we randomly select 50 Python repositories from our testing dataset, then for each GitHub repository}, we checkout all the TODO-introducing commits and extract their associated \textit{TODO-introduced methods}. 
Note that for building our dataset, we only use \textbf{paired} \textit{TODO-introduced methods} and excluded the \textbf{unpaired} ones. 
For this in-the-wild evaluation, both \textbf{paired} and \textbf{unpaired} \textit{TODO-introduced methods} are utilized. 
We then apply {\sc TDPatcher} to pinpoint if any methods' code blocks regarding this snapshot missed the introduced TODO comment. 
As a result, {\sc TDPatcher} reports 41 \textit{\major{TODO-missed} methods} in total, each \textit{\major{TODO-missed} method} and its original \textit{TODO-introduced method} pair are provided to three experts (all of which have Python programming experience for more than 5 years) for evaluation independently. 
Each expert manually labels the reported method as \textbf{True} or \textbf{False} or \textbf{Unsure} independently. 
The final results are determined by the majority rule of voting (if three different scores are obtained, the sample will be regarded as \textbf{Unsure}). 
Finally, after the manual checking, 26 methods are labeled as \textbf{True} (i.e., \textit{\major{TODO-missed} methods}), 10 methods are labeled as \textbf{False}, and 5 methods are labeled as \textbf{Unsure}. 
We have published the in-the-wild evaluation results in our replication package for verification~\cite{tdpatcher}.
% \footnote{\url{https://github.com/TDPatcher/TDPatcher}}. 
% after the manual checking, 
% we confirmed that \ref{} of them are \textit{TODO-dropped methods.}

We demonstrated three \textit{\major{TODO-missed} methods} detected by our {\sc TDPatcher} in Fig.~\ref{fig:in_the_wild_example}. 
The Ex.1 shows a \textbf{True} sample. 
In particular, after the developer ran the test case, several test methods failed (e.g., \texttt{test\_matcher\_match\_zero} and \texttt{test\_matcher\_match\_one\_plus}).  
In order to achieve short-term benefits (e.g., higher productivity or shorter release time), the developer marked these methods with the \texttt{xfail} marker (e.g., line 77/78, \texttt{@pytest.mark.xfail}). 
As a result, the details of these tests will not be printed even if they fail.
The developers added TODO comments here (e.g., Line 76/101) to remind themselves of these markers. 
Later, when they figured out the tests failed reasons, these markers can be removed. 
However, the same situation also applied to the method \texttt{test\_matcher\_phrase\_match}, but the corresponding TODO comment was forgotten to be added, which may cause the test method to be neglected. 

% Not all the reported \textit{TODO-dropped methods} can be manually confirmed.
The Ex.2 demonstrates an \textbf{Unsure} sample. 
The developer added a TODO comment (\textit{``TODO: - should check the original priority of the job''}) to \texttt{evluation\_finished} method. 
Our approach identified another \texttt{compilation\_finished} as a \textit{\major{TODO-missed} method} because of their similar local code implementations. 
Due to unfamiliarity with the code, it is challenging, even for human experts, to annotate such samples. 
For such cases, we still need developers to double-check the results. 
However, we argue that it is still beneficial to prompt the developers to be aware of such potential suboptimal places in time after they introduce any new TODOs.

Ex.3 demonstrates a common \textbf{False} sample in our in-the-wild evaluation. 
Given the \textit{TODO-introduced method} \texttt{esp\_test\_ts} with its added TODO comment (i.e., Line 66), {\sc TDPatcher} wrongly predicted this TODO comment should also be added to the \texttt{lyra\_test\_ts} method. 
The failed reason behind this sample is that the TODO comments within these two methods are slightly ``different'', while our approach can only handle methods missing the exactly \textbf{same} TODO comment. 
This sample also points out an interesting improvement for our approach, i.e., after patching the TODO comment to \textit{\major{TODO-missed} methods}, it is necessary to further  \textbf{update} the TODO comment according to a different method context, which could be handled by incorporating the comment updating tools~\cite{liu2020automating, lin2021automated, panthaplackel2020learning}. 
We will explore this research direction in our future work. 

\find{
{\bf How effective is {\sc TDPatcher} for detecting and patching \textit{\major{TODO-missed} methods} in the wild?}
{\sc TDPatcher} successfully detects 26 \textit{\major{TODO-missed} methods} from 50 Github repositories, which shows the effectiveness for detecting and patching \textit{\major{TODO-missed} methods} in real-world Github repositories. 
}

% \begin{figure}
% \vspace{0.0cm}
% \centerline{\includegraphics[width=0.49\textwidth]{pr-example.pdf}}
% % \vspace*{-5pt}
% \caption{Pull Request Examples}
% \label{fig:example}
% % \vspace{-5pt}
% \end{figure}

% \vspace{5pt}
% \noindent
% \framebox{\parbox{\dimexpr\linewidth-2\fboxsep-2\fboxrule}{
% \textbf{
% How effective is our {\sc TDCleaner} for removing obsolete TODO comments in real Github repositories? -- we conclude that our model is effective for detecting obsolete TODO comments in real word Github repos.}}} 
% \vspace{5pt} 

\section{Threats to Validity}
\label{sec:threats}
Several threats to validity are related to our research: 
% In the data preparation process, we removed the diffs which contain multiple TODOs. 
% When we evaluate our approach in practice, the evaluation data goes through the same data processing steps, which means the diffs with multiple TODOs are also removed before feeding them into our model. 
% We have to admit our current approach lacks the ability of dealing with multiple TODOs when analyzing unseen data. 
% We will try to address this shortcoming in our future work.

\vspace{0.1cm}\noindent\textbf{Internal Validity.}
Threats to internal validity relate to the potential errors in our code implementation and study settings. 
\major{Regarding the data preparation process, to ensure the quality of our constructed dataset, we only consider code snippets with the same TODO comments as equivalent suboptimal implementations, we removed TODO comments if they don't contain exactly the same words. 
In practice, similar TODO comments could also be associated with equivalent suboptimal implementations (as shown in Fig.~\ref{fig:in_the_wild_example} Ex.3).
These instances are missed by our data collection process. 
Detecting and adding such missed cases can enlarge our dataset and boost our model performance, we will focus on this direction in the future. 
We choose Python 2.7 and 3.7 standard library \texttt{ast} for parsing Python2 and Python3 projects, there may exist methods with new syntax that our parser cannot handle. Only 1.5\% (1,451 out of 97,413) methods failed to parse due to syntax errors, the potential threats of syntax errors are limited. 
Since we collected our dataset from top-10K Python projects which may contain noise, we further performed a manual validation, ensuring the quality of our constructed dataset. 
Regarding the evaluation settings, we evaluate our approach by extracting methods only from \textit{TODO-introduced} files, which is simpler than real-world situations. 
We consider this simplification as a trade-off to estimate our model's effectiveness and simulate real-world working environments, since more than 70\% \textit{TODO-introduced methods} are added within the same file. 
Expanding our approach to other files could introduce new challenges (i.e., data imbalance) and will be our future research direction.} 
In order to reduce errors in automatic evaluation, we have double-checked the source code, we have carefully tuned the parameters of the baseline approaches and used the highest-performing settings for each approach. 
Considering there may still exist errors that we did not note or neglect, we have published our source code to facilitate other researchers to replicate and extend our work. 

\vspace{0.1cm}\noindent\textbf{External Validity.}
Threats to external validity are concerned with the generalizability of our study. 
One of the external validities is the dataset, our dataset is constructed from Python projects in GitHub. 
This is because Python is one of the most popular programming languages widely used by developers in GitHub. 
Considering that our approach is language-independent, we believe that our approach can be easily adapted to other programming languages. 
Another external validity is that our work focused on TODO comment, which is a special case of Self-Admitted Technical Debt (SATD)~\cite{potdar2014exploratory}. 
Our preliminary study focuses on TODOs as opposed to other SATD (e.g., HACK, FIXME, NOTE). 
This is because TODO comments construct the major proportion of SATD. 
We will try to extend our work to other programming languages as well as other types of SATD to benefit more developers in the future. 

\vspace{0.1cm}\noindent\textbf{Model Validity.}
Threats to model validity relate to model structure that could affect the learning performance of our approach. 
In this study, we choose the state-of-the-art code embedding pre-trained model, GraphCodeBERT, for our approach. 
\major{GraphCodeBERT was pretrained on the CodeSearchNet dataset, which includes 2.3M functions of six programming languages paired with natural language documents. 
Our study focuses on methods related to TODOs, which means these methods' implementations are temporarily suboptimal. 
We believe that the potential threats of data leakage associated with employing GraphCodeBERT for our specific tasks are minimal. 
}
Recent research has proposed new models, such as codet5~\cite{wang-etal-2021-codet5}, copilot~\cite{chen2021evaluating} and alphacode~\cite{li2022competition} for different advanced programming tasks (e.g., code generation, competitive programming, etc).
Our approach does not shed light on the effectiveness of employing other advanced pre-trained models with respect to new structures and new features. 
We will explore other advanced models in future work.

% Regarding the model hyperparameters, there are two key hyperparameters for constructing our model, i.e., the embedding size of the encoders and the size of MLP hidden layers. 
% Because we use the pre-trained BERT model as our encoders, the embedding size is fixed to 768 for BERT. 
% Therefore, the only hyperparameter we can tune is the size of the hidden layers. 
% Theoretically, we can fine-tune the size of hidden layers, however, due to the wide range of different size settings, the complexity of training is too expensive and time-consuming. We thus follow classic MLP layer settings~\cite{gao2020technical} in previous works and our approach has achieved promising results for the task. 

% Another threat to validity is that our work is focused on TODO comment, which is a special case of Self-Admitted Technical Debt (SATD)~\cite{potdar2014exploratory}. 
% Our preliminary study focuses on TODOs as opposed to other SATD, such as TODO, HACK, FIXME, etc. This is because TODO comments constructs the major proportion of SATD. 
% We plan to investigate the effectiveness of our approach to other types of STAD comments in the future. 
% This is because a TODO often hints at the functionality that is not yet implemented, while HACK and FIXME might point to existing, but imperfect implementations. 
% Therefore, we choose the obsolete TODO comments for our preliminary study, we plan to investigate the effectiveness of our approach to other types of STAD comments in the future.

\section{Related Work}
\label{sec:related}
\subsection{TODO comments in software engineering}
TODO comments are extensively used by developers as reminders to describe the temporary code solutions that should be revisited in future. 
Despite the fact that TODO comments are widely adopted, the research works focus on TODO comments are still very limited. 
Previous studies have investigated the TODO comments usage in software development and maintenance~\cite{storey2008todo, sridhara2016automatically, nie2019framework, nie2018natural, gao2021automating}.

Storey et al.~\cite{storey2008todo} performed an empirical study among developers to investigate how TODO comments are used and managed in software engineering. 
They found that the use of task annotations varies from individuals to teams, and if incorrectly managed, they could negatively impact the maintenance of the system. 
Sridhara et al.~\cite{sridhara2016automatically} developed a technique to check the status of the TODO comments, their approach automatically checks if a TODO comment is up to date by using the information retrieval, linguistics and semantics methods. 
Nie et al.~\cite{nie2018natural} investigated several techniques based on natural language processing and program analysis techniques which have the potential to substantially simplify software maintenance and increase software reliability. 
Following that, they proposed a framework, TrigIt, to encode trigger action comments as executable statements~\cite{nie2019framework}. 
Experimental results show that TrigIt has the potential to enforce more discipline in writing and maintaining TODO comments in large code repositories.
Gao et al.~\cite{gao2021automating} proposed a deep-learning based approach TDCleaner to detect and remove obsolete TODO comments just-in-time by mining the commit histories of the software repositories. 

Different from the previous research, in this work we propose a novel task of \textit{\major{TODO-missed} methods} detection and patching. 
We build the first large dataset for this task and explore the possibility of automatically adding TODO comments to the \textit{\major{TODO-missed}} methods. 

\subsection{Self-admitted technical debt in software engineering}
TODO comment is a common type of self-admitted technical debt (SATD).
SATD was first proposed by Potdar and Shihab~\cite{potdar2014exploratory} in 2014, which refers to the technical debt (i.e., code that is incomplete, defective, temporary and/or suboptimal) consciously introduced by developers and documented in the form of comments they self-admit it. 
In recent years, various studies have investigated SATD from different aspects~\cite{sierra2019survey}, i.e., SATD detection, SATD comprehension and SATD repayment~\cite{collard2011lightweight, huang2018identifying, li2015systematic, ernst2015measure, potdar2014exploratory, da2017using, wehaibi2016examining, xavier2020beyond, zampetti2021self, muse2022fixme, zampetti2018self, zampetti2017recommending, maldonado2015detecting, melo2022identification, guo2021far, yu2020identifying, iammarino2021empirical, de2020identifying, phaithoon2021fixme, xiao2021characterizing, alhefdhi2023towards, mastropaolo2023towards, liu2020using, panthaplackel-etal-2020-learning}. 

\major{
Regarding SATD detection, Potdar et al.~\cite{collard2011lightweight} extracted 101,762 code comments from m 4 large open source systems, and manually identified 62 patterns that indicate SATD. 
Maldonado et al.~\cite{da2017using} presented an NLP-based method to automatically identify design and requirement self-admitted technical debt. 
Huang et al.~\cite{huang2018identifying} used text-mining based methods to predict whether a comment contains SATD or not. 
Yan et al.~\cite{yan2018automating} first proposed the change-level SATD determination model by extracting 25 change features, their model can determine whether or not a software change introduces TD when it is submitted. 
Regarding SATD comprehension, different studies have been conducted to understand the SATD in software development life cycles. 
For example, Maldonado et al.~\cite{maldonado2015detecting} manually analyzed 33,093 comments and classified them into 5 types (i.e., design debt, defect debt, documentation debt, requirement debt and test debt). 
Bavota et al.~\cite{bavota2016large} studied the introduced, removed, and unaddressed SATDs in projects' change history. 
Wehaibi et al.~\cite{wehaibi2016examining} empirically studied how technical debt relates to software quality from five large open-source software projects. 
Xiao et al.~\cite{xiao2021characterizing} conducted a qualitative analysis of 500 SATD comments in the Maven build system to better understand the characteristics of SATD in build systems. 
Regarding the SATD repayment, prior works study the removal of SATD from software systems, i.e., repaying the SATDs. 
Zampetti et al.~\cite{zampetti2018self} conducted an in-depth study on the removal of SATD. 
They found that 25\% to 60\% SATD were removed due to full class or method removal, and 33\% to 63\% SATD were removed as the partial change in their corresponding method. 
Most recently, Rungroj et al.~\cite{maipradit2020wait} introduced the concept of ``on-hold'' SATD and proposed a tool~\cite{maipradit2020automated} to identify and remove the ``on-hold'' SATD automatically. 
Mastropaolo et al.~\cite{mastropaolo2023towards} empirically investigated the extent to which technical debt can be automatically paid back by neural-based generative models. 
Alhefdhi et al.~\cite{alhefdhi2023towards} proposed a deep learning model, named DLRepay, for automated SATD repayment. 
}

Different from the aforementioned research, our work focuses on TODO-related methods. 
This is because the TODO comments are regarded as the most common type of SATD, which are widely used by developers across different software projects. 
Moreover, {\sc TDPatcher} can be easily transferred to other types of STAD (e.g., FIXME, HACK), we plan to adapt our model to other SATD in our future work. 
% Instead of detecting the SATDs within the existing software projects, our model first investigates the possibility of 
% we proposed a novel task of detecting and patching \textit{TODO-dropped methods}. 
% To the best of our knowledge, we are the first work to investigate the possibility of detecting \textit{TODO-dropped methods}.

\subsection{Code embeddings in software engineering}
Embedding is a technique from NLP (Natural Language Processing) that learns distributed vector representation for different entities (e.g., words, sentences). 
One of the typical embedding techniques is word2vec~\cite{mikolov2013efficient}, which encodes each word as a fixed-size vector, where similar words are close to each other in vector space. 
Recently, an attractive research direction in software engineering is to learn vector representations of source code~\cite{feng2020codebert, guo2020graphcodebert, wang-etal-2021-codet5, gao2020generating, gao2019smartembed, chen2021evaluating} via large pre-trained models. 

Feng et al.~\cite{feng2020codebert} proposed CodeBERT, which learns the vector representation from programming languages (PL) and natural language (NL) for downstream code-related tasks, e.g., code search and code documentation generation tasks. 
Different from the existing pre-trained models regard a code snippet as a sequence of tokens, 
Guo et al.~\cite{guo2020graphcodebert} presented GraphCodeBERT, which uses data flow in the pre-training stage and learns the inherent structure of code. 
The GraphCodeBERT has shown its effectiveness on four code-related tasks, including code search, clone detection, code translation, and code refinement. 
Wang et al.~\cite{wang-etal-2021-codet5} proposed CodeT5, which employs a unified framework to seamlessly support both code understanding and generation tasks (i.e., PL-NL, NL-PL, and PL-PL). 

In this research, we first adopt the GraphCodeBERT for the task of \major{TODO-missed} methods detection and patching, which provides contextualized embeddings for TODO comments and their associated suboptimal implementations.

\section{Conclusion and Future work}
\label{sec:con}
This research first propose the task of automatically detecting and patching \textit{\major{TODO-missed} methods} in software projects. 
To solve this task, we collect \textit{TODO-introduced methods} from the top-10,000 GitHub repositories. 
To the best of our knowledge, this is the first large dataset for \textit{\major{TODO-missed} methods} detection. 
We propose an approach named {\sc TDPatcher} (\textbf{\underline{T}}O\textbf{\underline{D}}O comment \textbf{\underline{Patcher}}), which leverage a neural network model to learn the semantic features and correlations between TODO comments and their suboptimal implementations. 
Extensive experiments on the real-world GitHub repositories have demonstrated effectiveness and promising performance of our {\sc TDPatcher}.
In the future, we plan to investigate the effectiveness of {\sc TDPatcher} with respect to other programming languages.
We also plan to adapt {\sc TDPatcher} to other types of task comments, such as FIXME, HACK.

\section{Acknowledgements}
\label{sec:ack}
% This research was partially supported by ARC Laureate Fellowship FL190100035, and National Research Foundation, Singapore, under its Industry Alignment Fund – Pre-positioning (IAF-PP) Funding Initiative. Any opinions, findings and conclusions or recommendations expressed in this material are those of the author(s) and do not reflect the views of National Research Foundation, Singapore. 
This research is supported by the Starry Night Science Fund of Zhejiang University Shanghai Institute for Advanced Study, Grant No. SN-ZJU-SIAS-001. 
This research is partially supported by the Shanghai Sailing Program (23YF1446900) and the National Science Foundation of China (No. 62202341). 
This research is partially supported by the Ningbo Natural Science Foundation (No. 2023J292)
The authors would like to thank the reviewers for their insightful and constructive feedback.

% \section{Automatic Evaluation}
% \label{sec:eval}
% \input{evaluation}

% \section{Human Evaluation}
% \label{sec:human_eval}
% \input{human_eval}

% \section{Practical Usage}
% \label{sec:practical}
% \input{practical}

% \section{Discussion}
% \label{sec:discussion}
% \input{disc}

% \section{Related Work}
% \label{sec:related}
% \input{related}

% \section{Conclusion and Future work}
% \label{sec:con}
% \input{conclusion}

% \section{Acknowledgements}
% \label{sec:ack}
% \input{acknowledgements}

\balance
\bibliographystyle{ACM-Reference-Format}
\bibliography{samples}

\end{document}